# Sewing skyrmion and antiskyrmion by quadrupole of Bloch points

Jin Tang[a,b*], Yaodong Wu[b], Jialiang Jiang[b], Lingyao Kong[a], Wei Liu[c*], Shouguo Wang[d], Mingliang Tian[a,b], and Haifeng Du[b*]

[a]School of Physics and Optoelectronic Engineering, Anhui University, Hefei, 230601, China

[b]Anhui Province Key Laboratory of Condensed Matter Physics at Extreme Conditions, High Magnetic Field Laboratory, HFIPS, Anhui, Chinese Academy of Sciences, Hefei, 230031, China

[c]Information Materials and Intelligent Sensing Laboratory of Anhui Province, Institutes of Physical Science and Information Technology, Anhui University, Hefei, 230601, China

[d]Anhui Key Laboratory of Magnetic Functional Materials and Devices, School of Materials Science and Engineering, Anhui University, Hefei 230601, China

*email: jintang@ahu.edu.cn; liuwei@ahu.edu.cn; duhf@hmfl.ac.cn

Magnetic skyrmions are particle-like spin swirls and are characterized by a topological charge, which is a quantized integer that is conserved under continuous deformation of the spin configuration [1]. Skyrmions have been proposed as potential information carriers in next-generation magnetic data storage and processing devices due to their emergent topology-related electron-magnetic properties [2]. Magnetic Bloch points are typically considered magnetization singularities, where magnetization vanishes [3-5]. In recent years, there has been increasing interest in the interaction between magnetic Bloch points and magnetic skyrmions [3-5]. Theory and simulations have shown the important role of Bloch points in stabilizing Y-shaped skyrmion strings [4], biskyrmions [3], and bobbers [5]. Understanding the interplay between magnetic Bloch points and skyrmions is therefore of great importance for the development of new materials and technologies based on these exotic magnetic structures [3-5].

Here, we show that the quadrupole of Bloch points can sew skyrmions and antiskyrmions in the depth dimension to form a new style of skyrmion-antiskyrmion coupling. Antiskyrmions are reported in few non-centrosymmetric $D_{2d}$ magnets [6, 7]. In $D_{2d}$ magnets, complex multiple magnetic interactions also contribute to the stabilization of elliptic Bloch-skyrmions [8]. The skyrmion-antiskyrmion coupling pair is not only a candidate for exploring fascinating particle-antiparticle interactions [9], but also a promising information carrier with zero skyrmion Hall effects for topological spintronic device applications [10]. However, in $D_{2d}$ magnets, skyrmion and antiskyrmion reveal repulsive interaction with each other in two dimensions [11].

In this communication, we report the direct connection between Bloch points, skyrmions, and antiskyrmions by extending the topological magnetism in the third dimension of a FeNiPdP alloy with $S_4$ symmetry [7], in a combination of 3D micromagnetic simulations with both Fresnel and electronic holography modes of Lorentz-transmission electronic microscopy (TEM) magnetic imaging [12]. Our results not only build complete physical models of topological spin textures in the popular antiskyrmion-hosting magnet FeNiPdP, but also propose a new skyrmion-antiskyrmion coupling style that is applicable in topological spintronic devices.

In 3D magnets, skyrmions form with spin twists along the depth orientation because of the magnetic dipole-dipole interactions or conical modulations [13, 14]. Using the MuMax3 approach, we simulated the 3D skyrmionic textures in chiral magnets with both $D_{2d}$ interactions and uniaxial magnetic anisotropies $K_u$. It has been established two types of skyrmionic textures in $D_{2d}$ magnets [8]: skyrmion and antiskyrmion (Fig. S1). In a 136-nm-thick magnet, our simulations show a 3D skyrmion string with a depth-modulated Bloch-to-Néel spin-twisting from the middle layer to the surface layer (Fig. S2a), which is attributed to the magnetic dipole-dipole interaction. Similarly, setting a uniform antiskyrmion string as the initial state, we obtain a stable twisted 3D antiskyrmion string (Figs. S2b and S3). A pair of magnetic singularities (Fig. S2b) supported by $D_{2d}$ interaction maintains the surface layers and contributes to the positive integer topological charge $Q = 1/(4\pi) \int\int \mathrm{d}^2\mathbf{r}\mathbf{m} \cdot$ $(\frac{\partial \mathbf{m}}{\partial x} \times \frac{\partial \mathbf{m}}{\partial y})$ , which essentially counts how many times the magnetization vector field $\mathbf{m}$ at the position $\mathbf{r} = (x, y)$ winds around the unit sphere [15]. However, the cross-sectional configurations along the locations of singularities are not energetically supported from the view of magnetic dipole-dipole interaction (Fig. S3c). Setting the skyrmion and antiskyrmion strings as two end states in the nudged elastic band (NEB) simulation [16], we obtain that their mutual transformation must be through two stable 3D hybrid strings mediated by the emergence and annihilation of Bloch points (Fig. 1c). First, through the emergence of a dipole of Bloch points, the skyrmion string transforms to a skyrmion-bubble string (Fig. S4) [17], whose surface magnetizations maintain the skyrmion with $Q = -1$ while interior magnetizations turn to a bubble with $Q = 0$. Second, the skyrmion-bubble string transforms to the skyrmion-antiskyrmion string, whose topological reversals are mediated by the quadrupole of Bloch points (Fig. 1a). Finally, the topological skyrmion-antiskyrmion string to antiskyrmion string transformation is achieved through the annihilation of the quadrupole of Bloch points. It should be noted that $Q$ is the product of vorticity and polarity (Fig. S1). Specifically, at negative magnetic fields, the $Q$ values for skyrmions and antiskyrmions are 1 and $-1$, respectively, due to the polarity reversal compared to the values at positive magnetic fields.

The simulations reveal that the skyrmion-antiskyrmion string is more stable than the antiskyrmion string from the view of total energy (Fig. 1c) and can be stabilized in a broad field region (Fig. S5). By fixing the thickness $t$, the exchange interaction stiffness $A$, and the saturated magnetization $M_s$, we obtain the stable phase diagram of the skyrmion-antiskyrmion string as a function of $K_u$ and $D_{2d}$, as shown in Fig. 1d. The skyrmion-antiskyrmion string can be stabilized in a broad region of magnetic parameters, and typically in chiral magnets with uniaxial anisotropies and relatively weak $D_{2d}$ interactions.

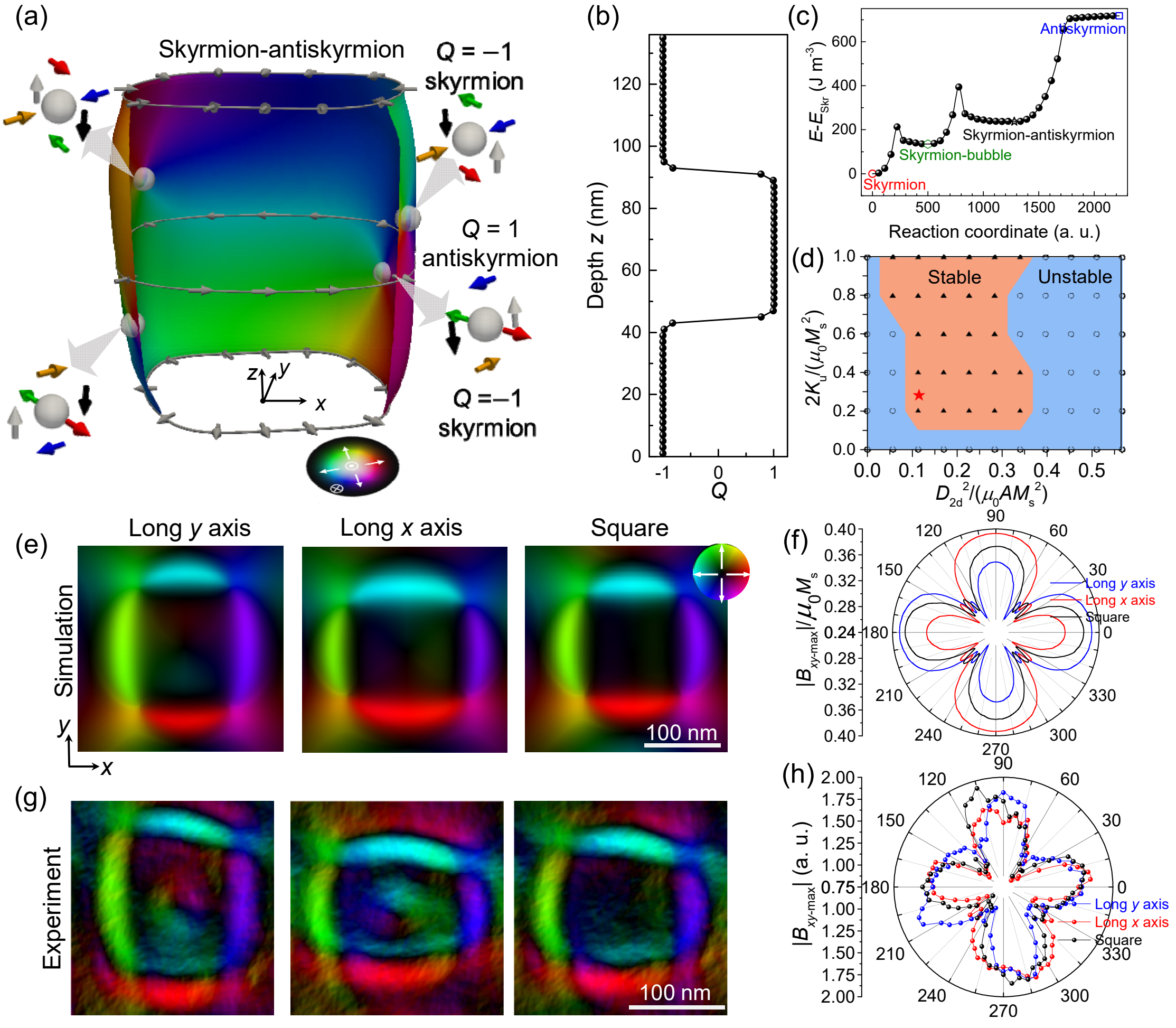

**Fig. 1.** (Color online) Theoretical prediction and experimental observation of the skyrmion-antiskyrmion string in $D_{2d}$ magnets with perpendicular magnetic anisotropies. (a) 3D configuration of skyrmion-antiskyrmion string. White arrows indicate the locations of the quadrupole of Bloch points. (b) Depth $z$ dependence of topological charge of the skyrmion-antiskyrmion string. (c) Topological transformation between the skyrmion string and the antiskyrmion string obtained using the NEB method. The energy is expressed by the differenced energy between each state and the skyrmion state $E-E_{Skr}$ (d) Stabilization of the 136-nm thick skyrmion-antiskyrmion string as a function of $D_{2d}$ and $K_u$. $B/(0.5\mu_0 M_s) = 0.46$. Here, $\mu_0$ is the vacuum permeability. Simulated (e) and experimental (g) in-plane average magnetic induction field $B_{xy}$ of skyrmion-antiskyrmion strings with different shapes. Simulated (f) and experimental (h) azimuthal angle $\theta$ dependence of normalized maximum in-plane magnetic induction field $|B_{xy\text{-max}}|$ for skyrmion-antiskyrmion strings. Magnetic parameters of FeNiPdP for simulations are marked by "★" in (d). The magnetic induction field mappings are retrieved from TIE analysis of Fresnel images taken at a defocused distance of 300 μm.

The measured magnetic parameters (marked by a star symbol "★" in Fig. 1d) of $(Fe_{0.67}Ni_{0.3}Pd_{0.07})_3P$ alloy with $S_4$ symmetry satisfy the stabilization region of the skyrmion-antiskyrmion string [7]. We synthesized the single crystal of $(Fe_{0.67}Ni_{0.3}Pd_{0.07})_3P$ using the self-flux method. Macroscopic magnetization measurements (Fig. S6 and S7) indicate that the magnetic parameters of our samples closely resemble those of the targeted sample [7]. Magnetic domains in 136-nm thick $(Fe_{0.67}Ni_{0.3}Pd_{0.07})_3P$ lamella are then experimentally observed using Lorenz-TEM, which images the average in-plane magnetic induction fields [12]. For consistency, we further obtain the simulated magnetic induction fields of 3D spin configurations (Figs. S8 and S9). Figs. 1e and 1g show the representative simulated and experimental in-plane magnetic induction field $B_{xy}$ mappings of skyrmion-antiskyrmion strings. The skyrmion-antiskyrmion strings can reveal various shapes, *i.e.*, the rectangle with a long *x*-axis, the rectangle with a long *y*-axis, and the square. Moreover, the in-plane average magnetization mappings of skyrmion-antiskyrmion strings with rectangle shapes show the mixed characteristics of skyrmions and antiskyrmions, *i.e.*, a weak Bloch-twisted skyrmion-like core encircled by twisted antiskyrmion-like configurations. Fresnel images of rectangle skyrmion-antiskyrmion strings also contain additional dotted contrasts in the center (Figs. S10 and S11). In contrast, Fresnel images of square skyrmion-antiskyrmion strings reveal no additional dotted contrasts in the center. When integrating the near-surface magnetic induction fields with $Q = -1$ of the rectangle skyrmion-antiskyrmion string, the Néel component cancels out but a weak residual Bloch-skyrmion-like configuration remains (Fig. S11). Similar phenomena have been observed and discussed in 3D dipolar skyrmions [14]. Further analysis shows that the outer twisted antiskyrmion-like configurations are contributed by antiskyrmions $Q = 1$ in the interior layers (Fig. S11). Thus, these distinct characteristics of the hybrid skyrmion-antiskyrmion configurations can be applied for experimental verifications of the 3D topological spin configurations with depth-modulated spin twists in $(Fe_{0.67}Ni_{0.3}Pd_{0.07})_3P$ alloys. After carefully examining our Lorentz-TEM studies, we identify the theory-predicted weak Bloch-twisted skyrmion-like cores of the 3D skyrmion-antiskyrmion configurations in experiments, as shown in Fig. 1g. In experiments, we also identify the over-defocused Fresnel images of the skyrmion-antiskyrmion strings with black and white dots (Fig. S10), corresponding to the counterclockwise and clockwise circulations of Bloch-twisted skyrmion-like cores, respectively. These experimental signatures (Fig. 1g) are highly reproducible and fully consistent with micromagnetic simulations (Fig. 1e), providing unambiguous experimental proof of these 3D hybrid spin configurations.

Notably, the in-plane magnetic induction field mappings of skyrmion-antiskyrmion strings are quite like that of antiskyrmion strings (Figs. S8 and S12) [7].

After carefully examining the detailed in-plane spin configurations, here we provide a protocol for distinguishing skyrmion-antiskyrmion strings from antiskyrmion strings using Lorentz-TEM. When integrating the in-plane magnetic induction fields in near-surface layers of the antiskyrmion string, the magnetic induction field mappings in the top and bottom surface layers with $Q$ = 1 both show two magnetic singularities, around which the in-plane magnetic induction fields in all layers through the depth all point to a same orientation (Fig. S12). Thus, the overall amplitude of the average in-plane magnetic induction field near the magnetic singularities in the four corners of the rectangle shows maximum values for antiskyrmion strings (Fig. S13). In contrast, for the skyrmion-antiskyrmion string, the surface magnetic induction field in the corner of the rectangle shows a reversed orientation as that in the interior layers (Fig. S11), leading to the greatly reduced average in-plane magnetic induction field in the corners. Here, we define $|B_{xy\text{-max}}|$ as the maximum amplitude of the average in-plane magnetic induction field in a line along the radial axis from the center (Fig. S13), $\theta$ is the angle between the radial axis and $+x$ axis. Thus, $|B_{xy\text{-max}}|$ as a function of $\theta$ could be taken as a protocol for distinguishing between the skyrmion-antiskyrmion and antiskyrmion strings. For the antiskyrmion string, the peak values of $|B_{xy\text{-max}}|$ are in the four corners of the rectangle (Fig. S13), *i.e.* $\theta \approx 45, 135, 225, 315^{\circ}$. In contrast, the peak values of $|B_{xy\text{-max}}|$ are in the central edges of the rectangle for the skyrmion-antiskyrmion string (Fig. 1f), *i.e.* $\theta = 0, 90, 180, 270^{\circ}$. Such a protocol works for all shapes of strings (Figs. 1f and S13c). We further extract the $\theta$ vs. $|B_{xy\text{-max}}|$ of representative spin configurations in experiments, as shown in Fig. 1h. We confirm the stabilization of skyrmion-antiskyrmion strings from the characteristics of the proposed protocol. Note that the magnetic field configurations retrieved from the TIE analysis of Fresnel images keep consistency for a defocused distance smaller than 300 μm (Fig. S14). Excessive defocusing in Fresnel imaging can introduce noticeable artificial distortions. Additionally, the average magnetic induction fields of skyrmion-antiskyrmion strings are also determined using electronic holography magnetic imaging (Fig. S15). Both the Fresnel and electronic holography modes provide the stabilization proof of skyrmion-antiskyrmion strings. Moreover, the antiskyrmion string is hardly visible in our samples (Fig. S16), which is understood by the instability of high-energy antiskyrmion strings from the NEB simulation (Fig. 1c).

Fig. 2a shows the stabilized phase diagram as a function of temperature $T$ and magnetic field $B$. In the $(Fe_{0.67}Ni_{0.3}Pd_{0.07})_3P$ lamella, we observe only stripe-to-skyrmion transformations in the $B$-increasing process. Magnetic configurations of skyrmion strings (Fig. S17) are consistent with previous studies [7]. Magnetic configurations in the $B$-decreasing process show different trends. At $T \leq 300$ K, the ferromagnet (FM) prefers to transform to skyrmion-bubble strings (Fig. 2b) in the $B$-decreasing process. Skyrmion-antiskyrmion strings are hardly observed at $T \leq 300$ K. In the high-temperature region near the Curie temperature $T_c \approx 380$ K, the skyrmion-antiskyrmion strings are the most stable phases in low-field regions from skyrmion-bubble strings in the $B$-decreasing process. Importantly, the topological transformations between skyrmion-antiskyrmion and skyrmion-bubble strings are reversible by varying magnetic fields (Fig. 2c), suggesting the controlled injection and annihilation of Bloch points. Using micromagnetic simulations, such transformations are understood from the energy landscape. The energy difference shows that skyrmion-antiskyrmion and skyrmion-bubble strings are the more stable phases at low field and high field, respectively (Fig. S18). In our zero-temperature simulations, we cannot directly achieve such transformations, which agrees with the absence of skyrmion-antiskyrmion strings at low temperatures in experiments. At high temperatures, the significant thermal fluctuation $E_{\mathrm{therm}}$ can assist a high-energy state in overcoming an energy barrier to achieve reversible topological transformations between skyrmion-antiskyrmion and skyrmion-bubble strings (Fig. 2c).

Despite that skyrmion-antiskyrmion strings are spontaneous low-energy states at high temperatures, we can also obtain metastable skyrmion-antiskyrmion strings at low temperatures through a field-cooling process as shown in Fig. S19. Increasing the field, the skyrmion-antiskyrmion string can transform to the skyrmion-bubble string (Figs. S19b and S20). But the reversed transformation in the $B$-decreasing process is not supported for $T < 320$ K. We finally obtain the stabilization diagram as a function of temperature and field in the $B$-increasing process from initial skyrmion-antiskyrmion strings at zero fields (Fig. S19c), which shows that the skyrmion-antiskyrmion strings can stabilize at a broad temperature and field region as a metastable phase.

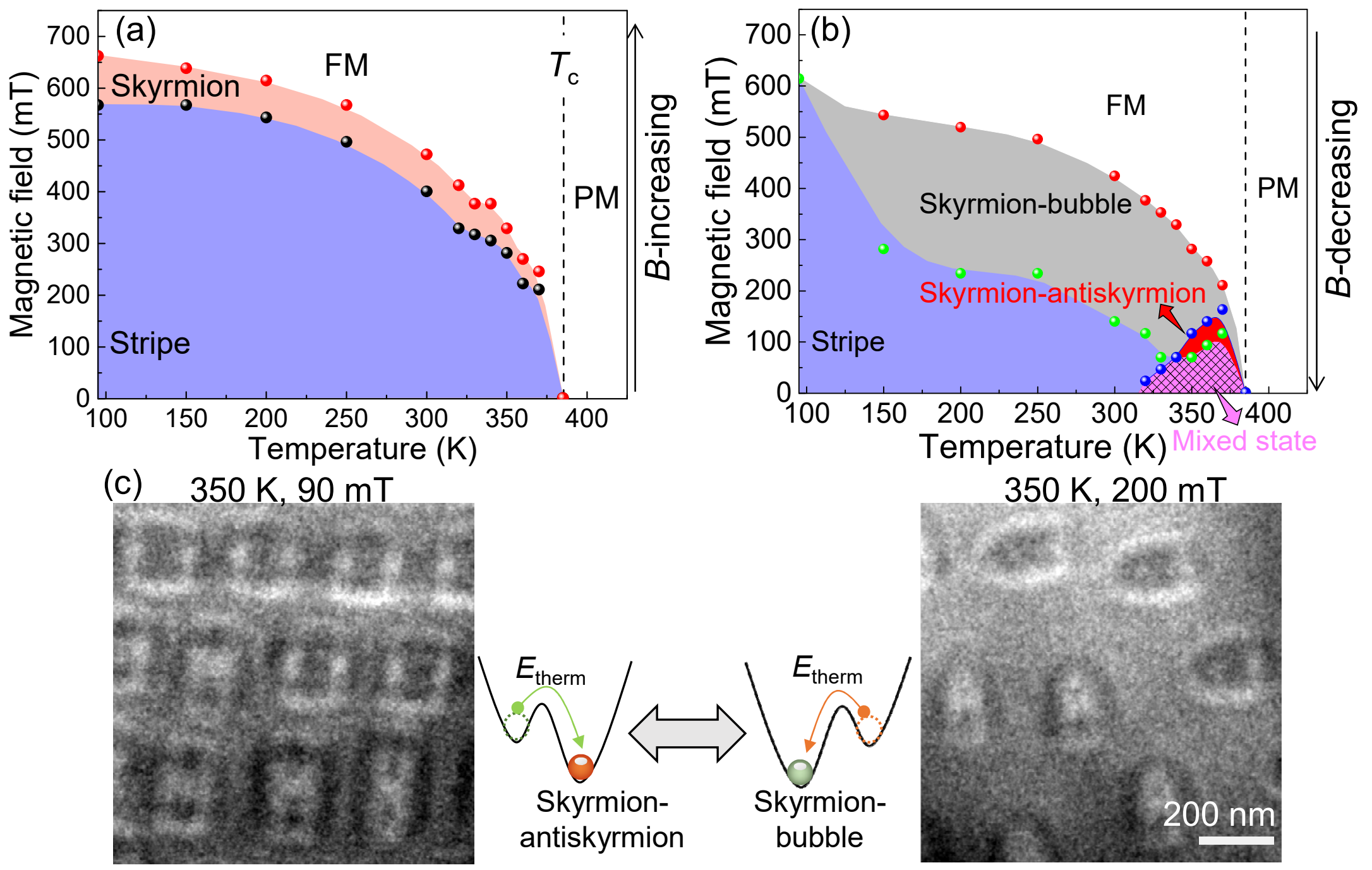

**Fig. 2**. (Color online) (a) Magnetic phase diagram in the field-increasing process. We increased the field step by step with a field interval of 20 mT. (b) Magnetic phase diagram in the field-decreasing process. We decreased the field step by step with a field interval of 20 mT. Here, mixed state means the coexisted skyrmion-antiskyrmion strings and stripe domains. (c) Reversible transformation between skyrmion-antiskyrmion and skyrmion-bubble strings by varying magnetic field at $T = 350$ K. The dots in (a) and (b) represent threshold magnetic fields between different phases. Defocused distance, −300 μm.

Skyrmion-antiskyrmion strings have similar overall in-plane magnetizations as that of antiskyrmion strings [7]. Thus, without consideration of 3D spin twists, skyrmion-antiskyrmion strings could be easily mistaken as antiskyrmion strings observed by Lorentz-TEM. Despite the similarity in overall average in-plane magnetic induction field mappings and Fresnel contrasts for antiskyrmion and skyrmion-antiskyrmion strings (Figs. 1e and S13a), they have entirely different topologies and related topological magnetism (Fig. 1), such as skyrmion Hall effects [18]. The skyrmion-antiskyrmion string with a greatly decreased averaged $Q$ could provide a new strategy to realize Hall balance (see detail in Supplemental Section II)

[10].

Thus, the observation of the skyrmionic textures and physical clarifications of magnetic nature in these popular antiskyrmion-hosting materials should redefine their potential applications in topological spintronic devices. Simulations based on measured magnetic parameters of $(Fe_{0.67}Ni_{0.3}Pd_{0.07})_3P$ show the lower energy of the skyrmion-antiskyrmion string and negligible energy barrier from the antiskyrmion string to the skyrmion-antiskyrmion string (Fig. 1c), which suggests that the skyrmion-antiskyrmion string could be achieved much easier than the antiskyrmion string in $(Fe_{0.67}Ni_{0.3}Pd_{0.07})_3P$ and experimentally identified in our experiment. Nevertheless, despite our ability to relax equilibrium antiskyrmion strings from initial configurations of antiskyrmions in zero-temperature simulations (Fig. S2b), the energy barrier for the transformation from antiskyrmion strings to skyrmion-bubble strings is too small to be reached in our NEB simulation. As a result, thermal fluctuations in realistic experimental conditions can readily lead to the annihilation of antiskyrmion strings. This explains why we consistently observed skyrmion-antiskyrmion strings instead of antiskyrmion strings (Fig. S16).

In summary, in combination with 3D micromagnetic simulations, we have demonstrated the stabilization and observation of 3D topological spin textures including skyrmion-antiskyrmion strings in chiral materials with both $D_{2d}$ and $K_u$. We propose a protocol to distinguish between skyrmion-antiskyrmion and antiskyrmion strings using Lorentz-TEM. Our results suggest a strategy to design skyrmion-antiskyrmion coupled states through the emergent quadrupole of Bloch points, clarify 3D spin configurations of topological spin textures in the popular antiskyrmion-hosting magnets, and provide more chances to develop potential topological spintronic devices.

**Acknowledgments**

This work was supported by the National Key R&D Program of China, Grant No. 2022YFA1403603; the Natural Science Foundation of China, Grants No. 12174396, 12104123, 1197402, 12204006, and 12241406; the National Natural Science Funds for Distinguished Young Scholar, Grants No. 52325105; Anhui Provincial Natural Science Foundation, Grants No. 2308085Y32 and 2108085QA21; Natural Science Project of Colleges and Universities in Anhui Province, Grant No. 2022AH030011; the Strategic Priority Research Program of Chinese Academy of Sciences, Grant No. XDB33030100; CAS Project for Young Scientists in Basic Research, Grant No. YSBR-084; and Systematic Fundamental Research Program Leveraging Major Scientific and Technological Infrastructure, Chinese Academy of Sciences, Grant No. JZHKYPT-2021-08.

**Author contributions**

Mingliang Tian, Jin Tang, and Haifeng Du supervised the project. Jin Tang conceived the idea and designed the experiments. Wei Liu synthesized the FeNiPdP single crystals. Jin Tang fabricated nanostructures and performed the TEM measurements with the help of Jialiang Jiang and Yaodong Wu. Jin Tang and Lingyao

Kong performed the simulations. Jin Tang and Haifeng Du wrote the manuscript with input from all authors. All authors discussed the results and contributed to the manuscript.

## Appendix A. Supplementary materials

Supplementary materials to this short communication can be found online.

## Note

During the reviewing process, we became aware of independent work supporting our conclusions [19].

# Supporting Information

## Sewing skyrmion and antiskyrmion by quadrupole of Bloch points

Jin Tang[a,b*], Yaodong Wu[b], Jialiang Jiang[b], Lingyao Kong[a], Wei Liu[c*], Shouguo Wang[d], Mingliang Tian[a,b], and Haifeng Du[b*]

[a]School of Physics and Optoelectronics Engineering, Anhui University, Hefei, 230601, China

[b]Anhui Province Key Laboratory of Condensed Matter Physics at Extreme Conditions, High Magnetic Field Laboratory, HFIPS, Anhui, Chinese Academy of Sciences, Hefei, 230031, China

[c]Information Materials and Intelligent Sensing Laboratory of Anhui Province, Institutes of Physical Science and Information Technology, Anhui University, Hefei, 230601, China

[d]School of Materials Science and Engineering, Anhui University, Hefei 230601, China

*email: jintang@ahu.edu.cn; liuwei@ahu.edu.cn; duhf@hmfl.ac.cn

**Supplementary Section I**

**Methods**

*Micromagnetic simulations*: The zero-temperature micromagnetic simulations were performed using MuMax3.[1] We consider the Hamiltonian exchange interaction energy ($\varepsilon_{\mathrm{ex}}$), 2D chiral DMI energy ($\varepsilon_{\mathrm{D2d}}$), uniaxial magnetic anisotropy energy ($\varepsilon_{\mathrm{u}}$), Zeeman energy ($\varepsilon_{\mathrm{zeeman}}$), and magnetic dipole-dipole interaction energy ($\varepsilon_{\mathrm{dem}}$) [1]. The total energy terms are given by:

$$\varepsilon = \int_{V_s}\{\varepsilon_{\mathrm{ex}} + \varepsilon_{\mathrm{u}} + \varepsilon_{\mathrm{zeeman}} + \varepsilon_{\mathrm{dem}} + \varepsilon_{\mathrm{D2d}}\}\mathrm{d}\boldsymbol{r} \quad \text{(S1)}$$

Here, exchange energy $\varepsilon_{\mathrm{ex}} = A(\partial_x\mathbf{m}^2 + \partial_y\mathbf{m}^2 + \partial_z\mathbf{m}^2)$, $\varepsilon_{\mathrm{D2d}} = D_{2\mathrm{d}}\mathbf{m}\cdot(y\partial_y\mathbf{m} - x\partial_x\mathbf{m})$, uniaxial magnetic anisotropy energy $\varepsilon_{\mathrm{u}} = -K_{\mathrm{u}}(\mathbf{m}\cdot\mathbf{n}_{\mathrm{u}})^{\mathbf{2}}$, Zeeman energy $\varepsilon_{\mathrm{zeeman}} = -M_{\mathrm{s}}\mathbf{B}_{\mathbf{ext}}\cdot\mathbf{m}$, and demagnetization energy $\varepsilon_{\mathrm{dem}} = -\frac{1}{2}M_{\mathrm{s}}\mathbf{B}_{\mathrm{d}}\cdot\mathbf{m}$. Here, $\mathbf{m}\equiv\mathbf{m}(x,y,z)$ is the normalized units continuous vector field that represents the magnetization $\mathbf{M}\equiv M_{\mathrm{s}}\mathbf{m}(x,y,z)$. The parameters $A$, $D_{2\mathrm{d}}$, $K_{\mathrm{u}}$, and $M_{\mathrm{s}}$ are the exchange interaction, DMI, uniaxial anisotropy constant, and saturation magnetization, respectively. $\mathbf{n}_{\mathrm{u}}$ is the unit vector field of the uniaxial easy magnetization axis and is set along the (001) $z$-axis. $\mathbf{B}_{\mathrm{d}}$ is the demagnetization field.

The exchange stiffness is fixed as $A = 8.1$ pJ · m$^{-1}$, and the saturated magnetization is fixed as $M_{\mathrm{s}} = 417$ kA · m$^{-1}$. We modify $K_{\mathrm{u}}$ and $D_{2\mathrm{d}}$ to obtain the phase diagram of stability for the skyrmion-antiskyrmion string using the conjugate-gradient method. The star symbol "★" in Fig. 1d represents the material parameter of FeNiPdP ($K_{\mathrm{u}} = 31$ kJ/m$^3$, $D_{2\mathrm{d}} = 0.2$ mJ/m$^3$) [2]. The cell size was set at $2\times2\times2$ nm$^3$.

The NEB calculations are performed with JuMag package, for detail about the NEB method in JuMag (https://github.com/ww1g11/JuMag.jl). In the NEB calculations, the calculation system, mesh size, and parameters are chosen to be the same as those in the Mumax3 simulations. We simulated the Lorenz-TEM images using a modified MALTS tool [3].

A Zhang-Li spin-transfer torque is considered for simulating current-driven dynamics [4]. The LLG equation including the spin transfer torques is written as:

$$\gamma \mathbf{m} \times \mathbf{H}_{\mathrm{eff}} = [(\mathbf{u} + \mathbf{v}) \cdot \nabla]\mathbf{m} - \mathbf{m} \times [(\beta\, \mathbf{u} + \alpha\, \mathbf{v}) \cdot \nabla]\mathbf{m} \quad \text{(S2)}$$

Here, $\alpha$ is the Gilbert damping and $\beta$ is the non-adiabatic parameter. $\mathbf{H}_{\mathrm{eff}}$ is the total effective field. The strength of spin transfer torque is characterized by $\mathbf{u} = \frac{gP\mu_B}{2eM_s}\mathbf{j}$ with parameters $g$, $\mu_B$, $e$, and $P$ are the Landé factor, Bohr magneton, electron charge, and the polarization rate, respectively. $\mathbf{v}$ is the drift velocity.

*Bulk sample preparations:* Single FeNiPdP crystals with $S_4$ symmetry were grown by the self-flux method with stoichiometric iron (Alfa Aesar, >99.9%) and Nicole (Alfa Aesar, >99.9%), palladium (Alfa Aesar, >99.9%), and red phosphorus (Alfa Aesar, >99.9%). The sintered bulk FeNiPdP was obtained by heating the mixture at 1100°C for 4 days. The crystal quality and structure group were both verified using Cu $K_\alpha$ radiation (TTR3 diffractometer, Rigaku).

*Fabrication of FeNiPdP microdevices*: The 136-nm thick FeNiPdP films were fabricated from a bulk single crystal using a standard lift-out method, with a focused ion beam and scanning electron microscopy dual beam system (Helios Nanolab 600i, FEI).

*TEM measurements*: We used both in-situ Fresnel and electronic holography imaging in Lorentz-TEM (Talos F200X, FEI) with an acceleration voltage of 200 kV to investigate magnetic domains in the FeNiPdP nanostructure. The TEM holder (model 636.6, Gatan) can support the temperature range from 95 to 350 K.

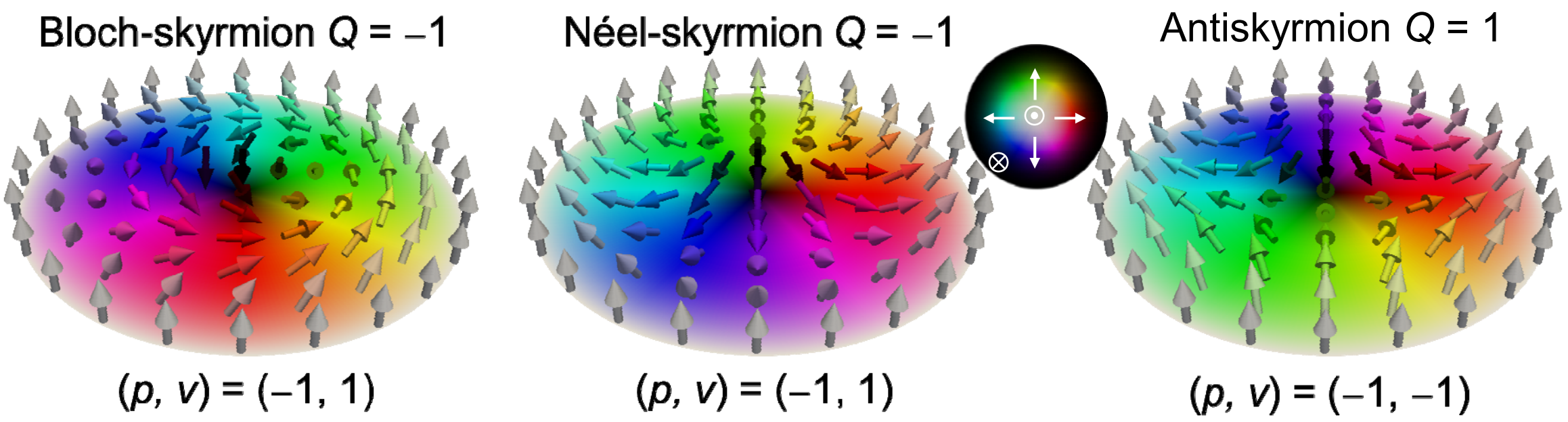

Fig. S1. Representative configurations of the 2D Bloch-skyrmion, the Néel-skyrmion, and the antiskyrmion. $p$ is the polarity, $v$ is the vorticity, and topological charge $Q = pv$.

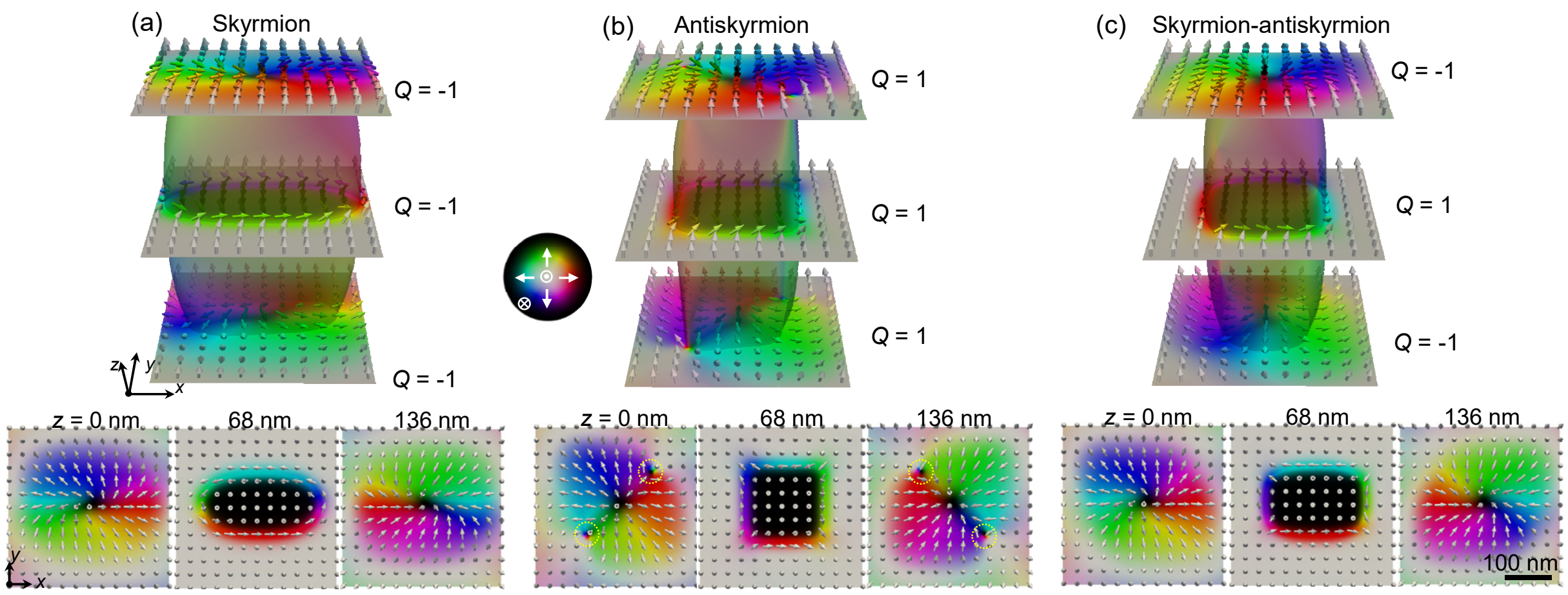

Fig. S2. Simulated 3D magnetic configurations of (a) the skyrmion string, (b) the antiskyrmion string, and (c) the skyrmion-antiskyrmion string. The dashed circles in b mark the locations of magnetic singularities. Magnetic parameters are marked by "★" in Fig. 1(d).

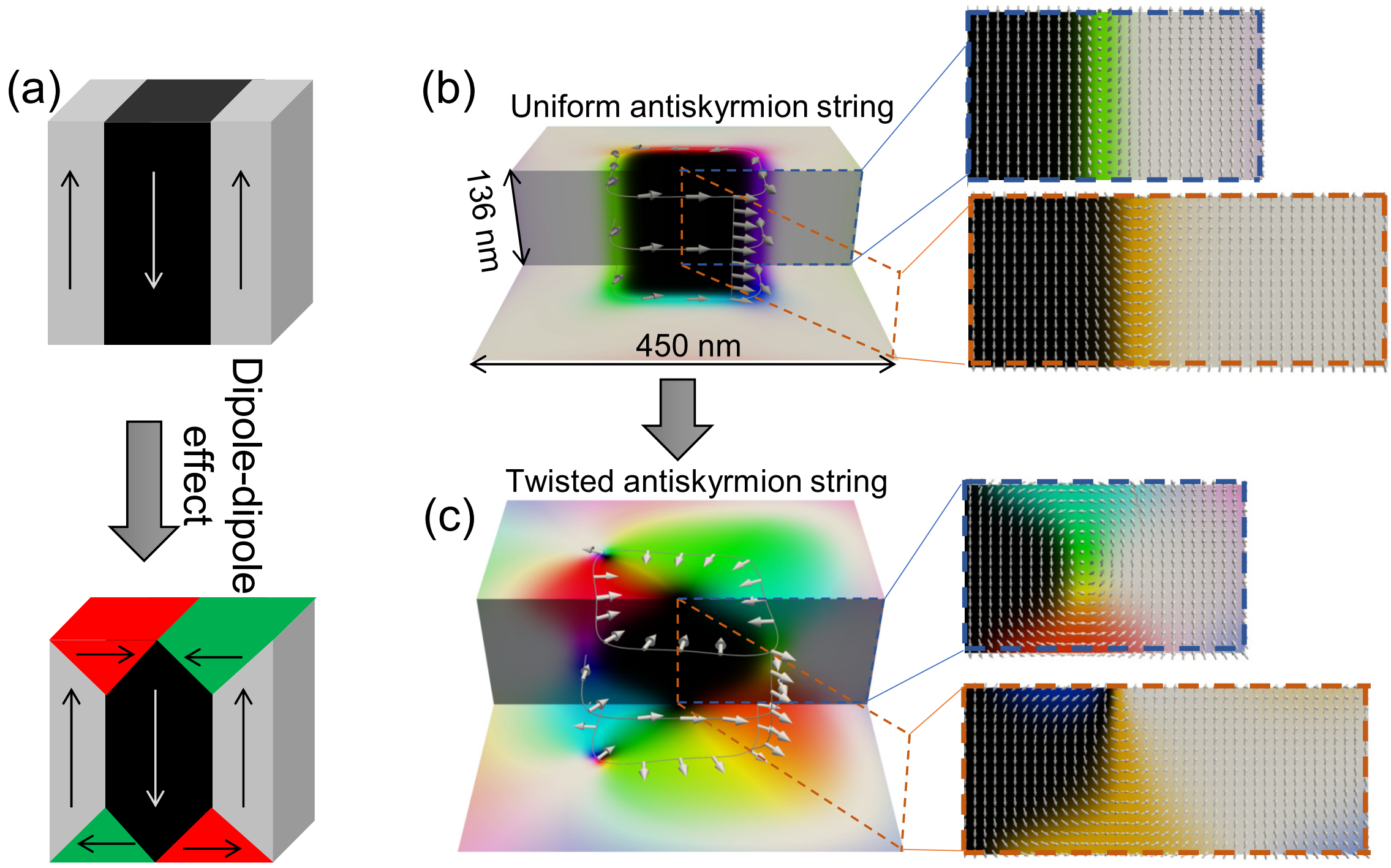

Fig. S3. (a) Schematic depth-modulated spin-twisting induced by the magnetic dipole-dipole interaction. (b) Uniform antiskyrmion string and cross-sectional configurations without magnetic dipole-dipole interaction. (c) Twisted antiskyrmion string and cross-sectional configurations induced by the magnetic dipole-dipole interaction. $B/(0.5\mu_0 M_s) = 0.46$.

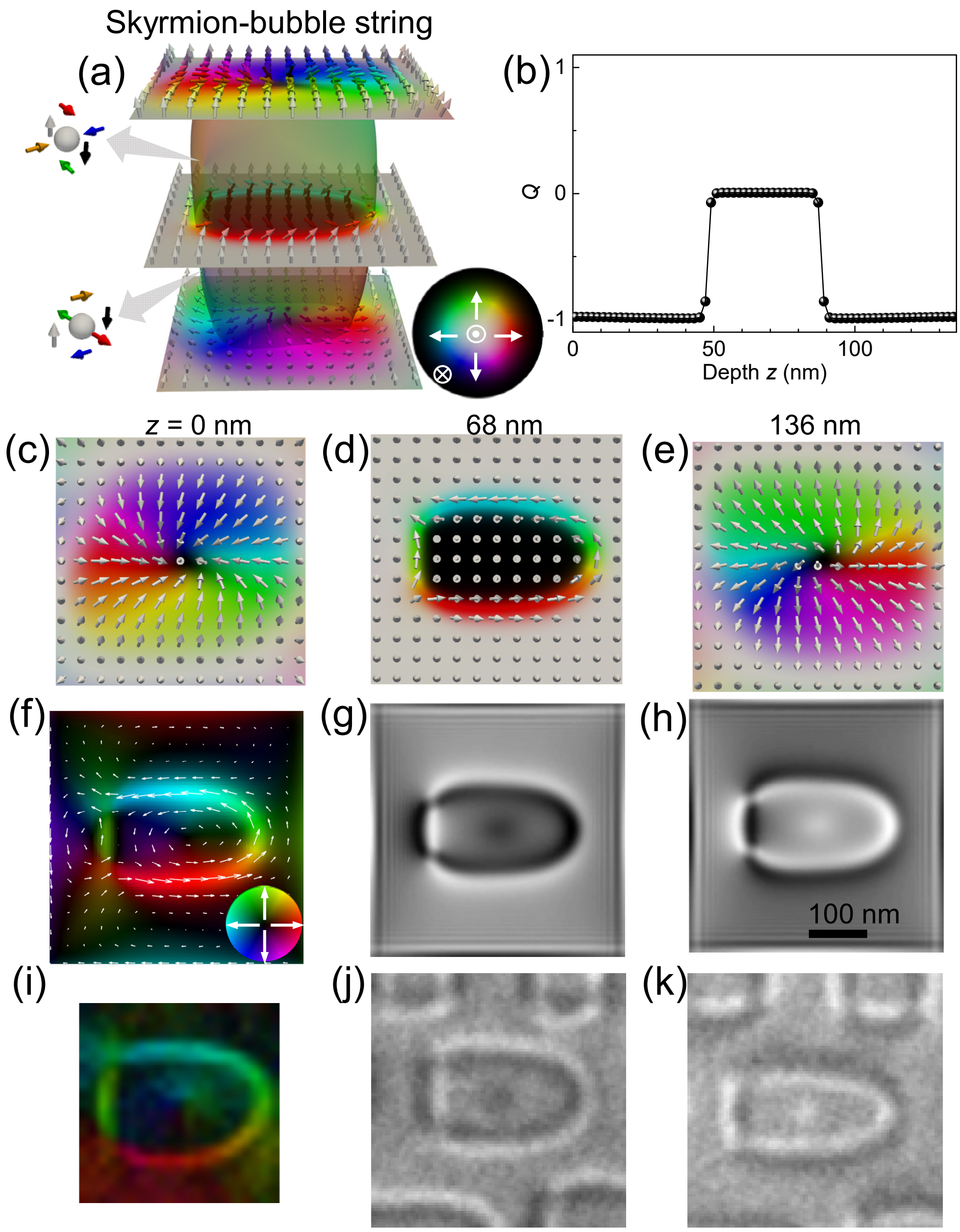

Fig. S4. (a) Simulated 3D configurations of the skyrmion-bubble string. (b) Depth $z$ dependence of the topological charge $Q$. (c)-(e) Magnetic configurations $m_{xy}$ at (c) the bottom layer $z$ = 0 nm, (d) the middle layer $z$ = 68 nm, and (e) the top layer $z$ = 136 nm. (f) Overall average in-plane magnetic induction field mapping $B_{xy}$. (g)-(h) Simulated Fresneal images at the (g) under-defocused and (h) over-defocused conditions. $B/(0.5\mu_0 M_s)$ = 0.46. (i) Experimental magnetic induction field $B_{xy}$ of the skyrmion-bubble string obtained from electronic holography technique. (j) and (k) Experimental Fresneal images at the (j) under-defocused and (k) over-defocused conditions at $B$ = 200 mT. Defocued distance, +300 μm.

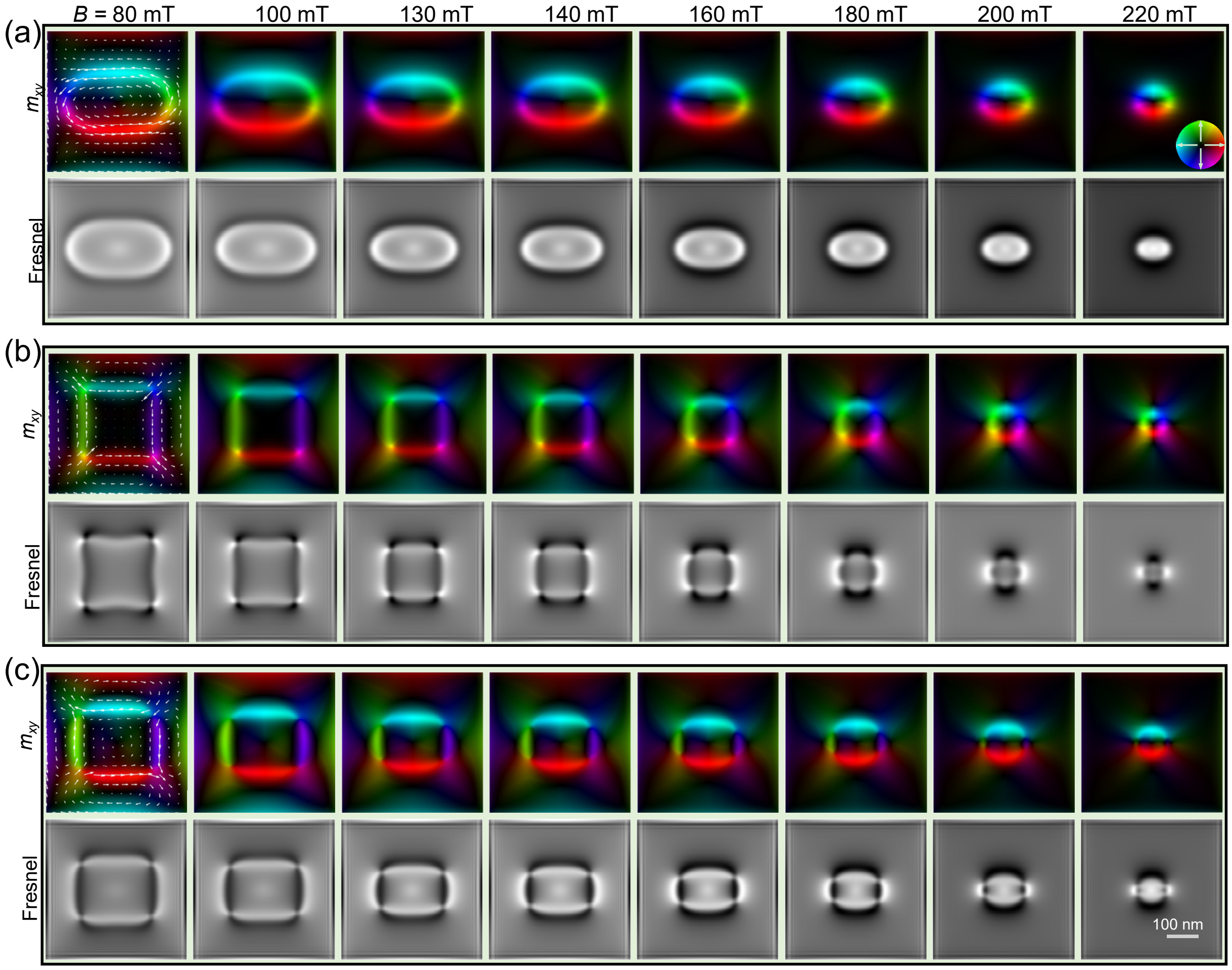

Fig. S5. Simulated magnetic field dependence of in-plane magnetization ($m_{xy}$) mappings and corresponding over-defocused Fresnel images of (a) skyrmion strings, (b) antiskyrmion strings, and (c) skyrmion-antiskyrmion strings.

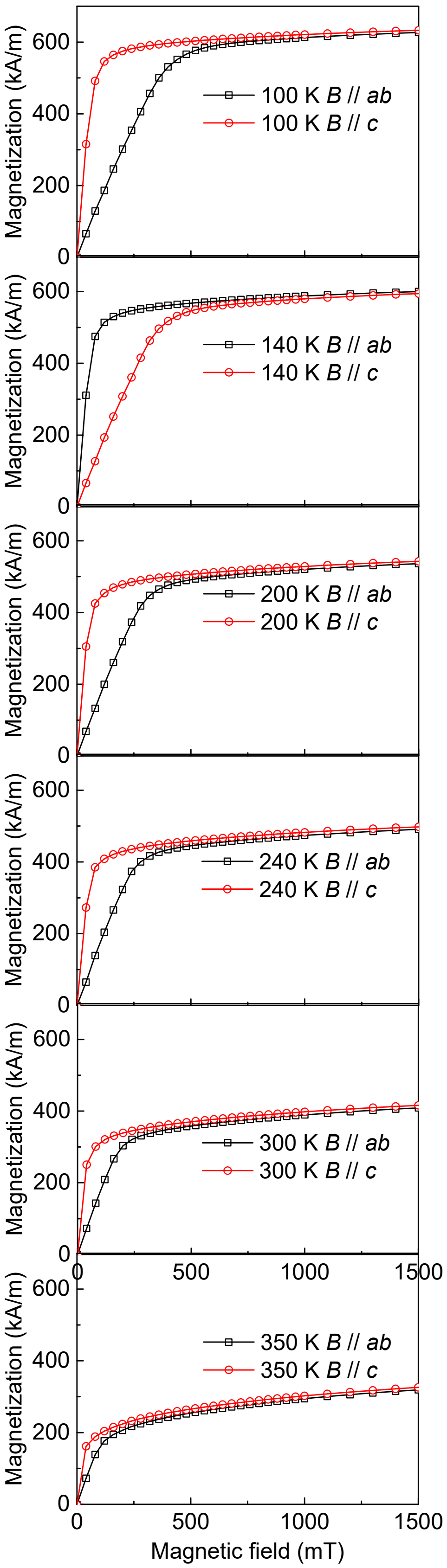

Fig. S6. Magnetization as a function of the magnetic field along the *c*-axis and *ab*-plane measured on a cube-shaped bulk FeNiPdP magnet.

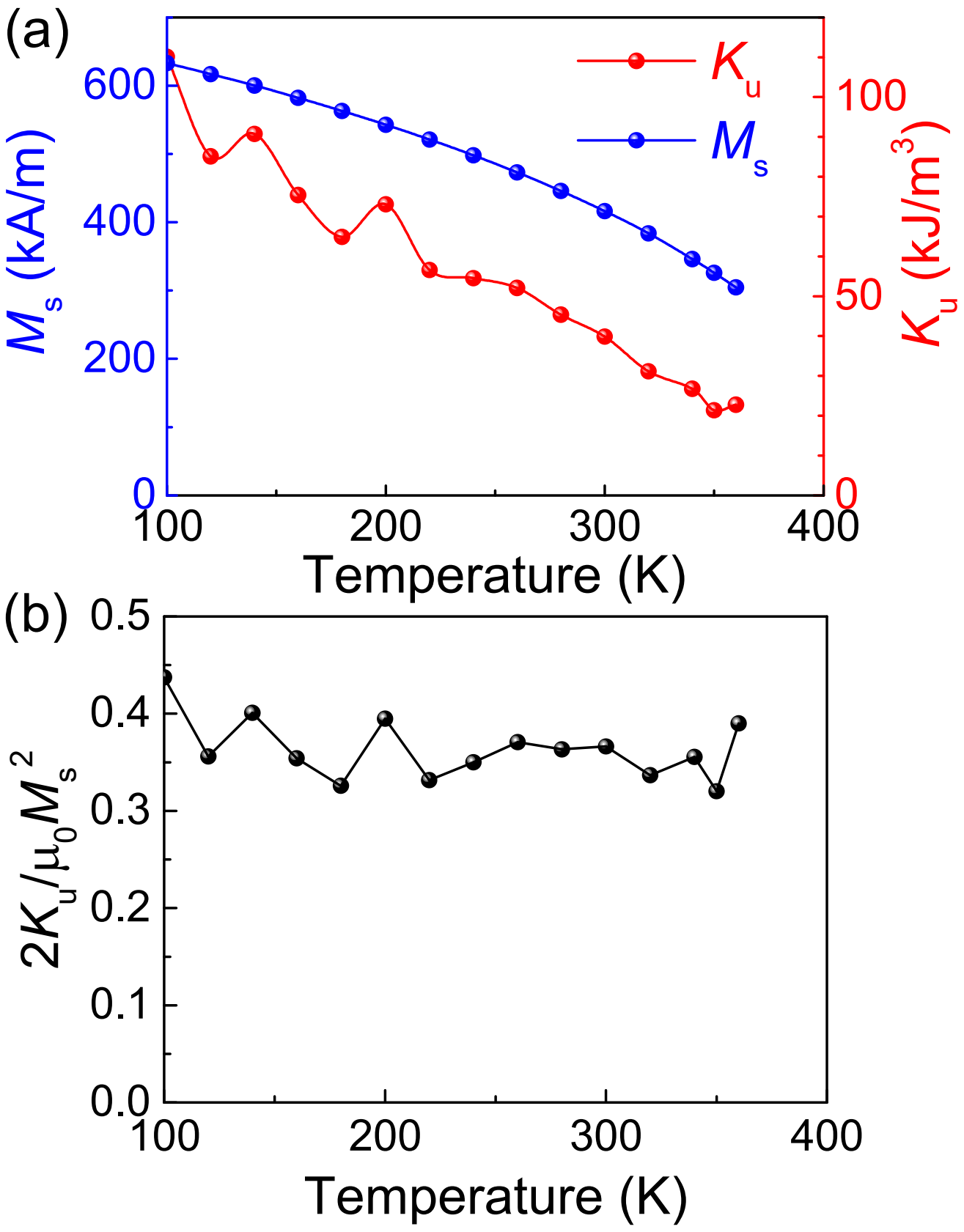

Fig. S7. (a) Temperature dependence of saturated magnetization $M_s$ and uniaxial magnetic anisotropy $K_u$. $M_s$ is taken as the magnetization along the *c*-axis at 1.5 T. The uniaxial magnetic anisotropy ($K_u$) is calculated from the differential values of $\int_0^{B_{Sat}} M(B)\mathrm{d}B$ between along and perpendicular to the *c*-axis. (b) Temperature dependence of quality factor $\frac{2K_u}{u_0 {M_s}^2}$.

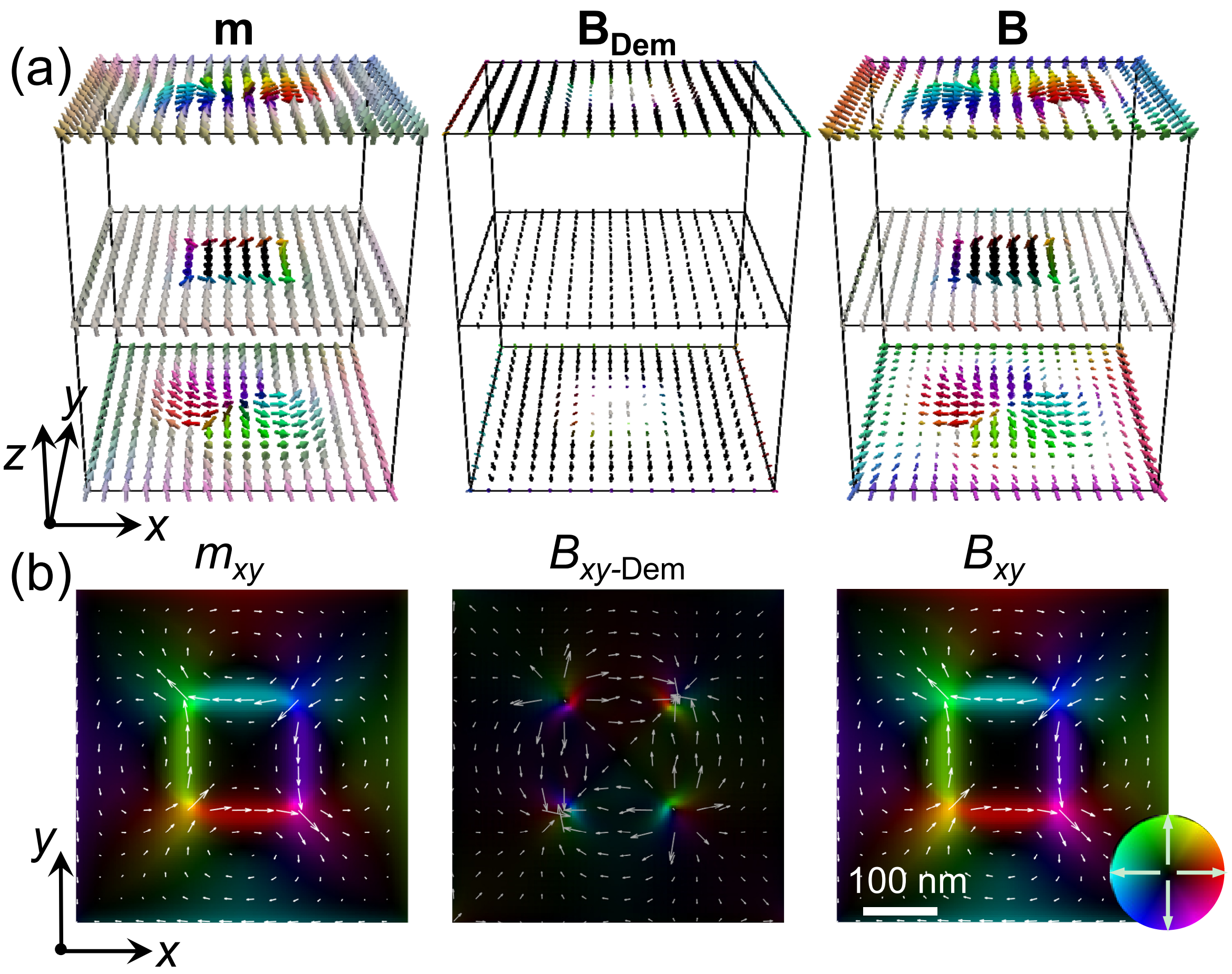

Fig. S8. (a) Simulated 3D configurations and (b) average in-plane of magnetization **m**, demagnetization field $\mathbf{B}_{\mathbf{Dem}}$, and total magnetic induction field **B** of the antiskyrmion string.

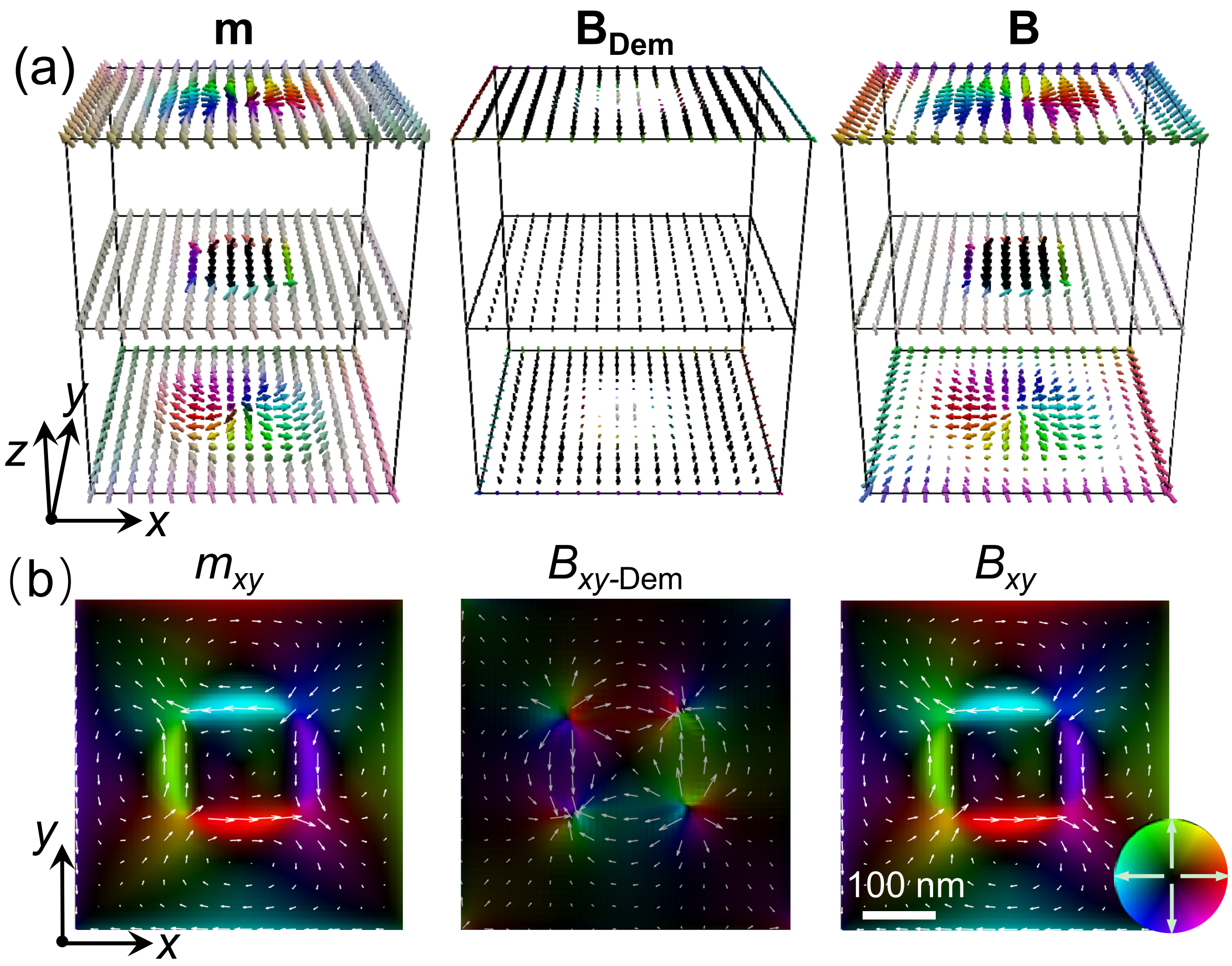

Fig. S9. (a) Simulated 3D configurations and (b) average in-plane of magnetization **m**, demagnetization field $\mathbf{B}_{\mathbf{Dem}}$, and total magnetic induction field **B** of the antiskyrmion-antiskyrmion string.

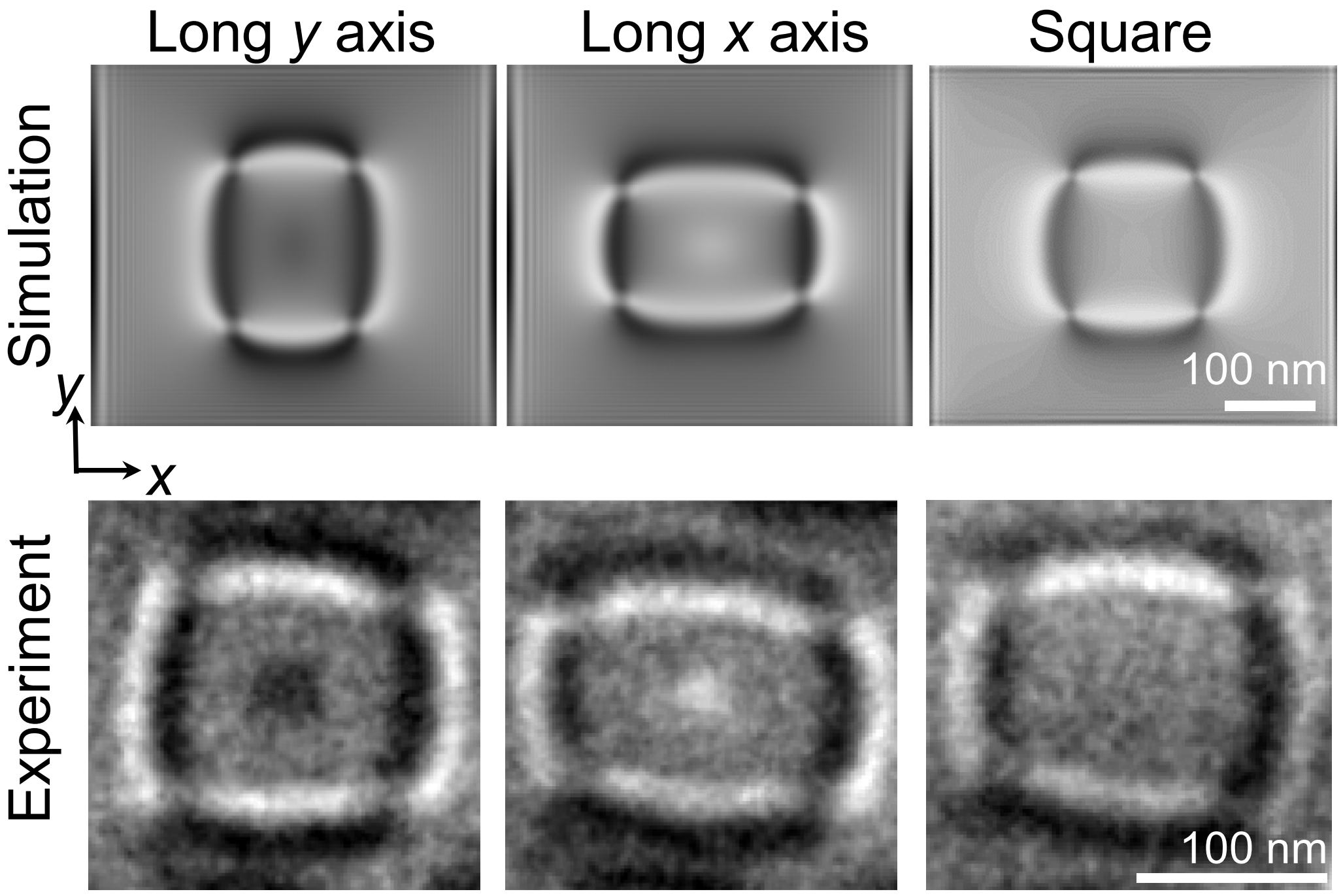

Fig. S10. Simulated and experimental Fresnel contrasts of skyrmion-antiskyrmion strings with different shapes. Defocused distance, 300 μm.

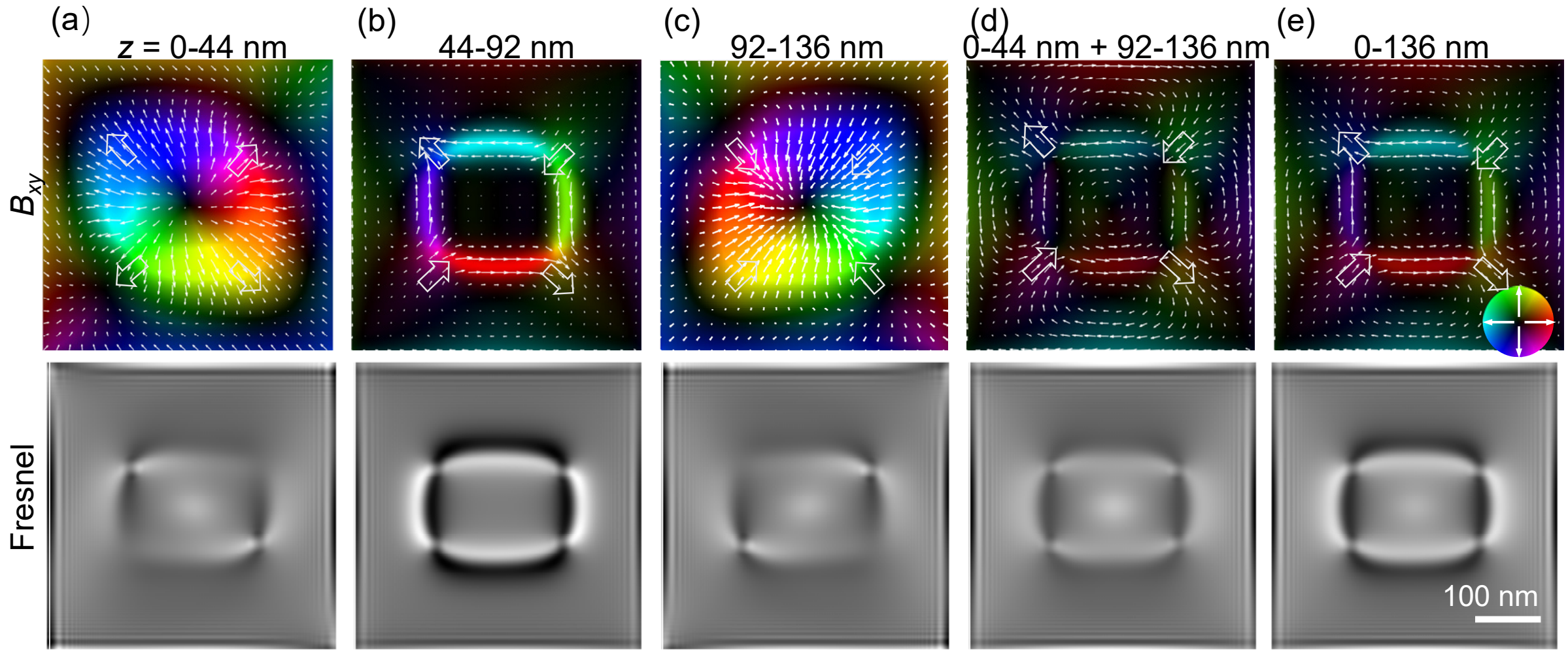

Fig. S11. Simulated in-plane magnetic induction mappings ($B_{xy}$) and corresponding over-defocused Fresnel images of the 3D skyrmion-antiskyrmion string for magnetizations in (a) the bottom surface layers $z = 0 – 44$ nm, (b) the interior layers $z = 44 – 92$ nm, (c) the top surface layers $z$ = 92-136 nm, (d) the near-surface layers $z = 0 – 44$ nm and 92- 136 nm, and (e) overall layers $z$ = 0-136 nm.

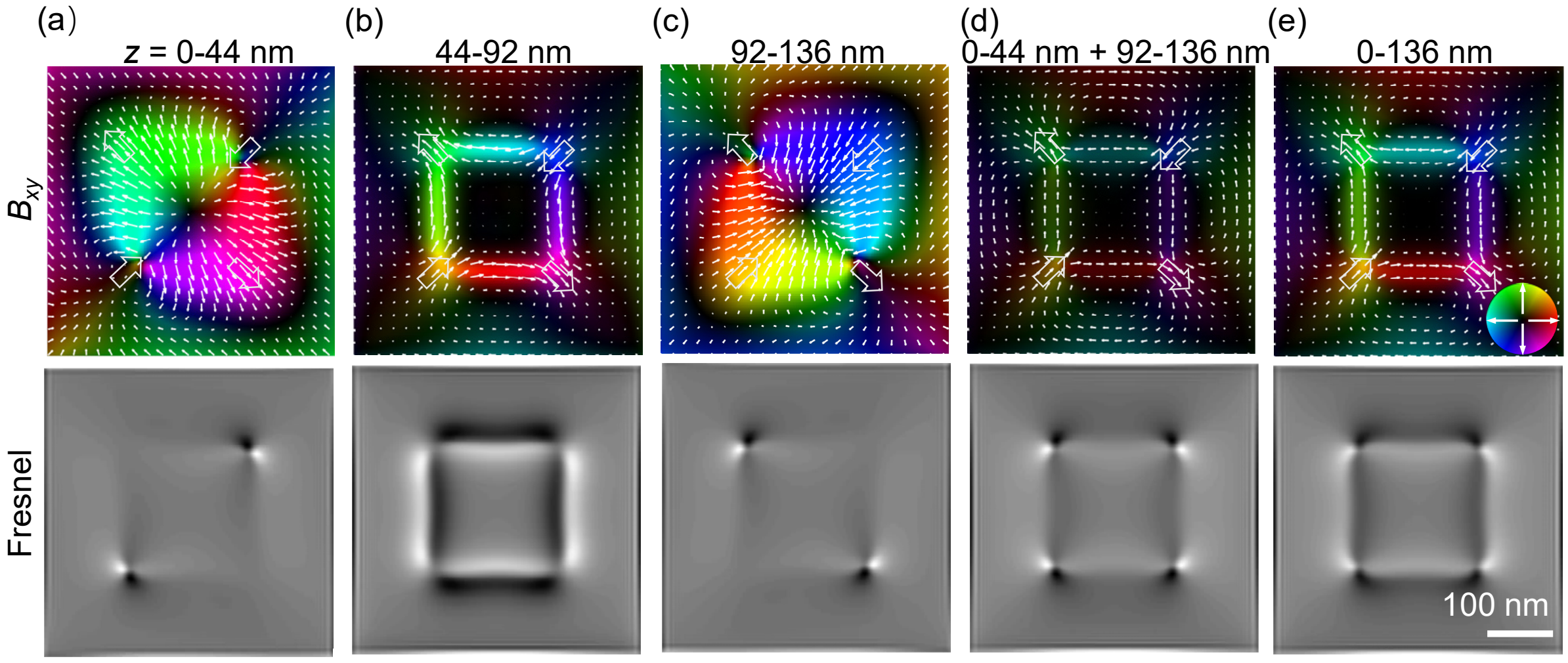

Fig. S12. Simulated in-plane magnetic induction field mappings ($B_{xy}$) and corresponding over-defocused Fresnel images of the 3D antiskyrmion string for magnetizations in (a) the bottom surface layers $z = 0 - 44$ nm, (b) the interior layers $z = 44 - 92$ nm, (c) the top surface layers $z = 92–136$ nm, (d) the near-surface layers $z = 0 - 44$ nm and 92- 136 nm, and (e) overall layers $z = 0–136$ nm. The spin orientation at the locations of magnetic singularities all point toward the same direction through the layers.

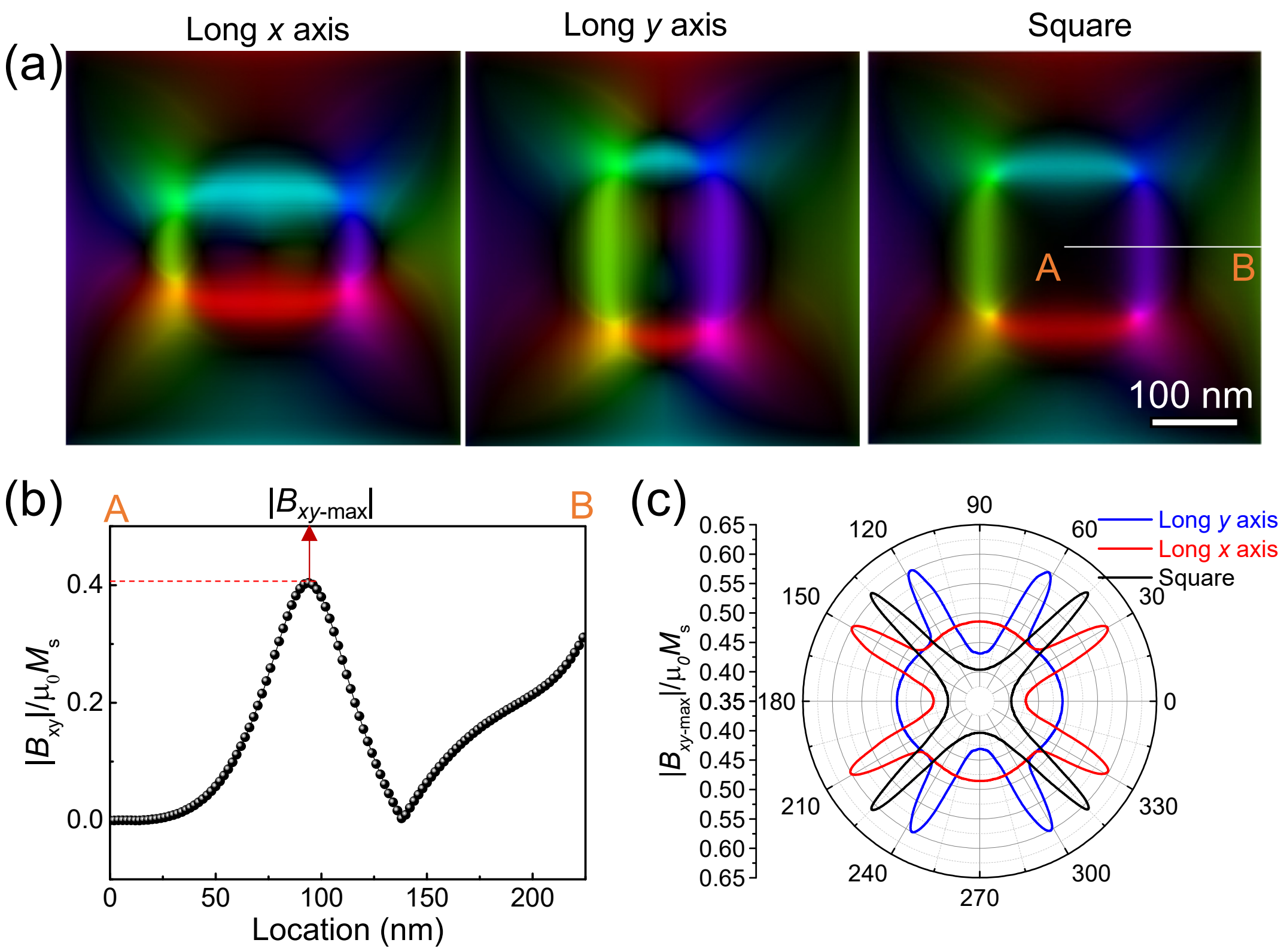

Fig. S13. Simulated average in-plane magnetic induction field mappings ($B_{xy}$) of antiskyrmion strings with different shapes. (b) Location dependence of the average normalized amplitude of in-plane magnetic induction field |$B_{xy}$| along the line AB shown in (a). (c) Simulated azimuthal angle $\theta$ dependence of maximum normalized in-plane magnetization induction field |$B_{xy\text{-max}}$| for antiskyrmion strings.

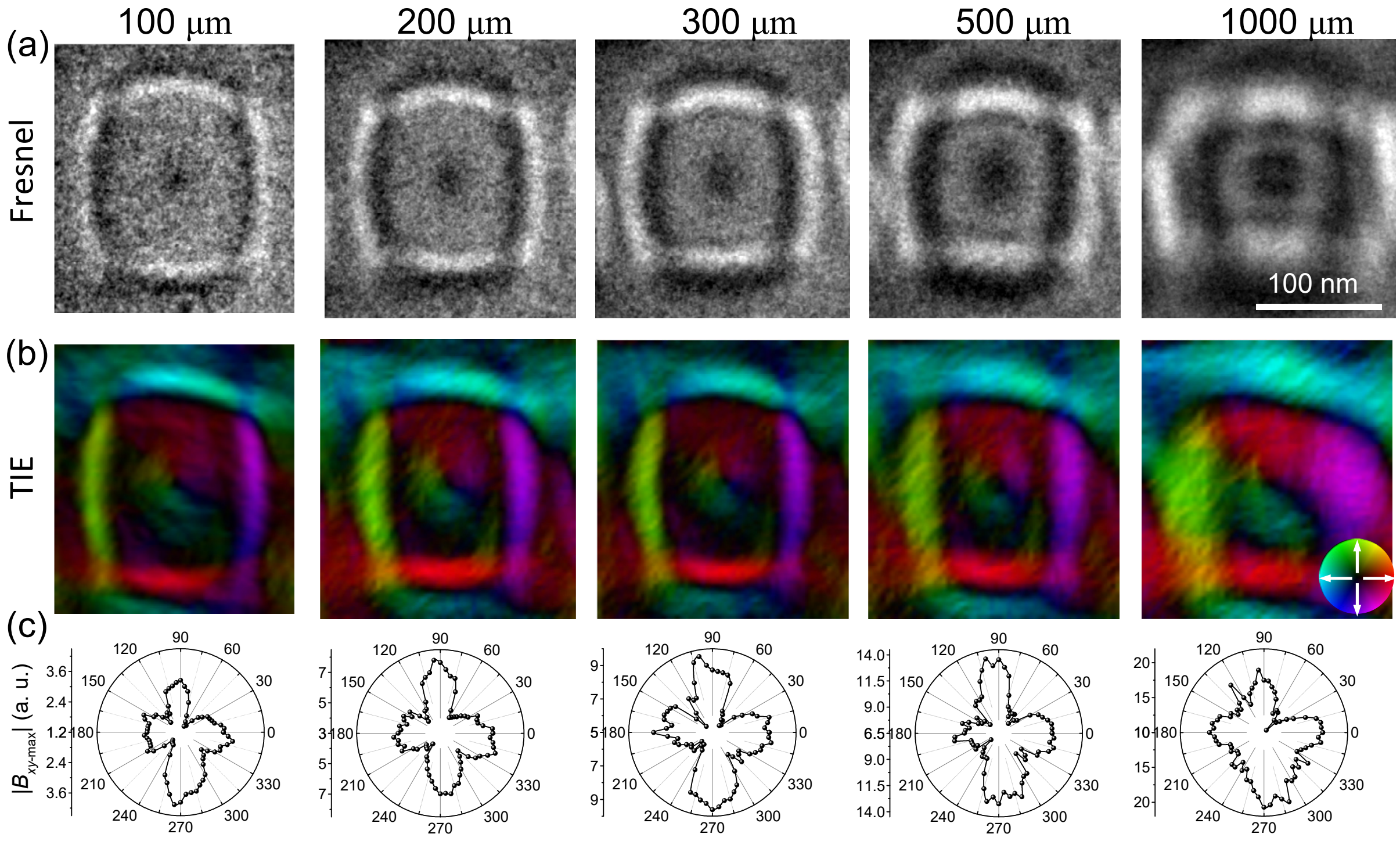

Fig. S14. (a) Over-defocused Fresnel images of skyrmion-antiskyrmion strings. (b) Retrieved magnetic induction field mappings using TIE analysis. (c) Angular dependence of the amplitude of maximum magnetic induction field $|B_{xy\text{-max}}|$. External magnetic field $B$ = 100 mT.

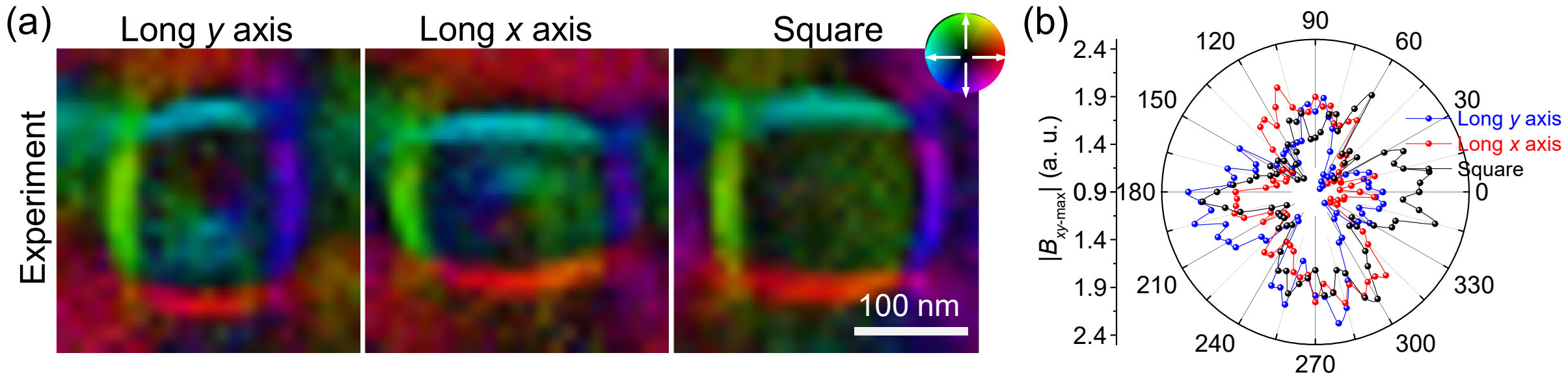

Fig. S15. (a) Retrieved magnetic induction field mappings of skyrmion-antiskyrmion strings using electronic holography. (b) Angular dependence of the amplitude of maximum magnetic induction field $|B_{xy\text{-max}}|$. External magnetic field $B$ = 180 mT.

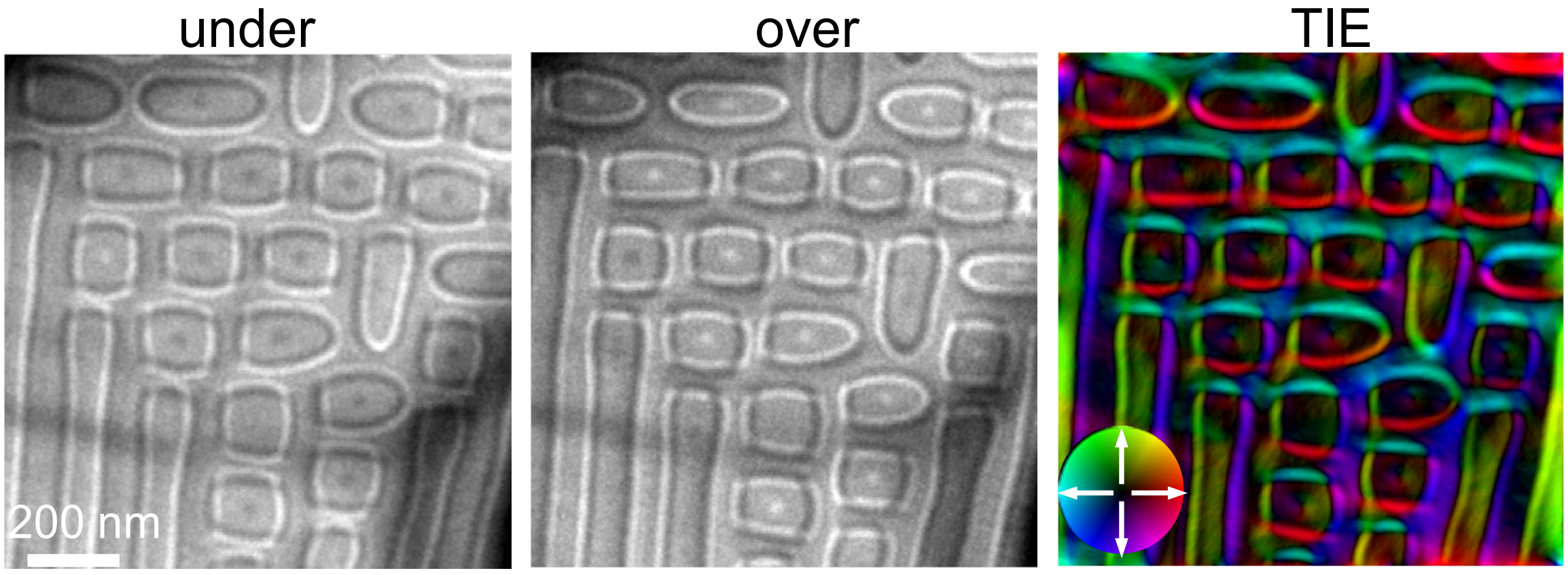

Fig. S16. Zero-field hybrid skyrmion-antiskyrmion states at temperature $T$ = 95 K. (a) under-defocused, (b) over-defocused, and (c) TIE reconstruction of in-plane magnetic induction field mappings. Defocused distance, 300 μm. No antiskyrmion strings can be obtained. $B$ = 100 mT.

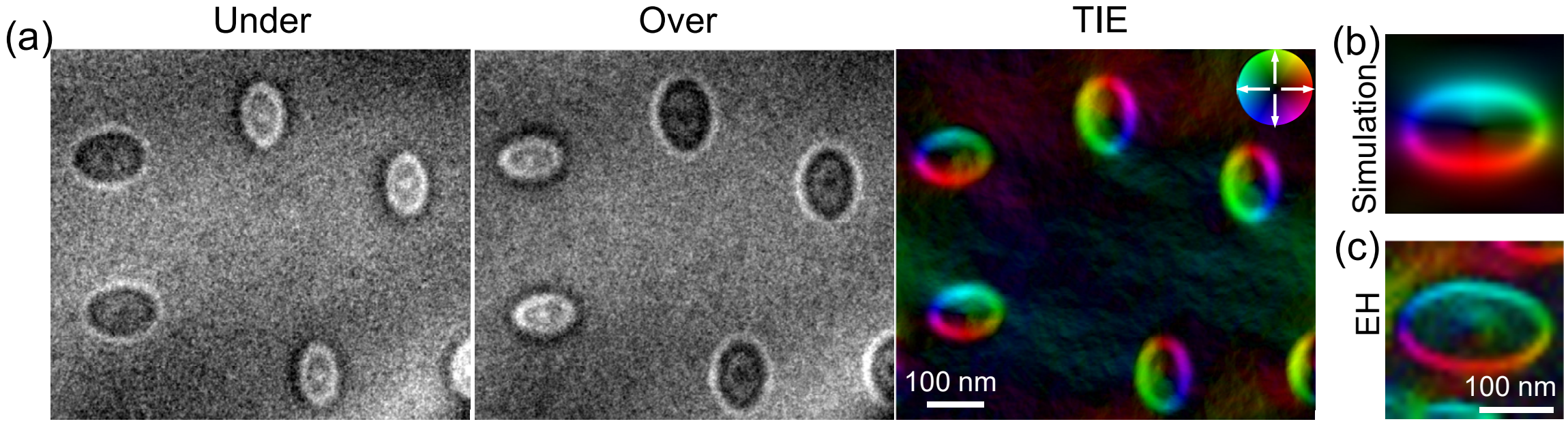

Fig. S17. (a) Experimental magnetic induction field mapping of skyrmion strings retrieved from TIE analysis of Fresnel images taken at a defocused distance of 200 μm. External magnetic field $B$ = 600 mT. (b) Simulated magnetic induction field of the skyrmion string. (c) Experimental magnetic induction field mapping of skyrmion strings retrieved from electronic holography (EH) technique. External magnetic field $B$ = 180 mT.

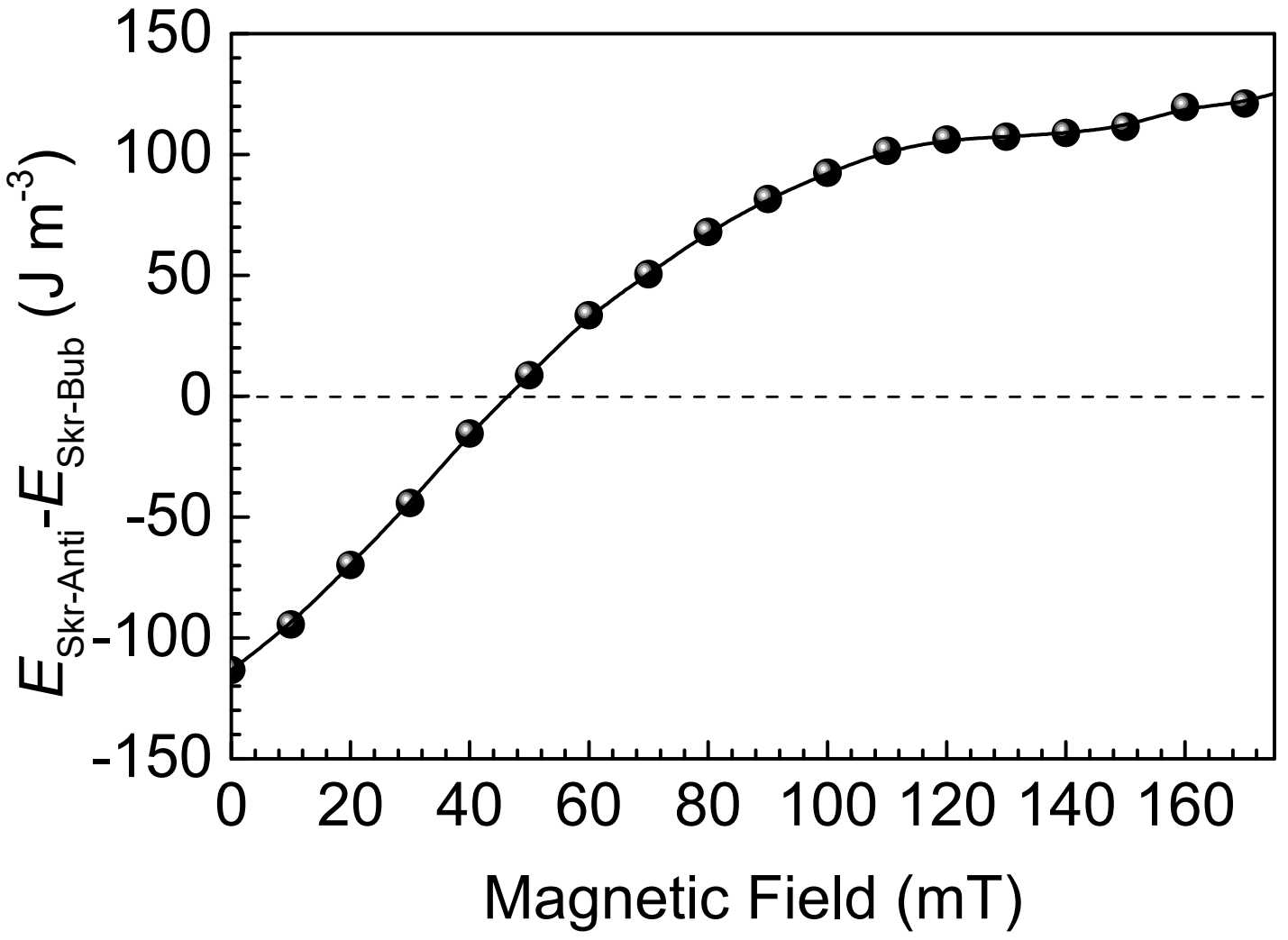

Fig. S18. Simulated magnetic field dependence of energy difference between skyrmion-antiskyrmion and skyrmion-bubble strings $E_{\text{Skr-Anti}} - E_{\text{Skr-Bub}}$.

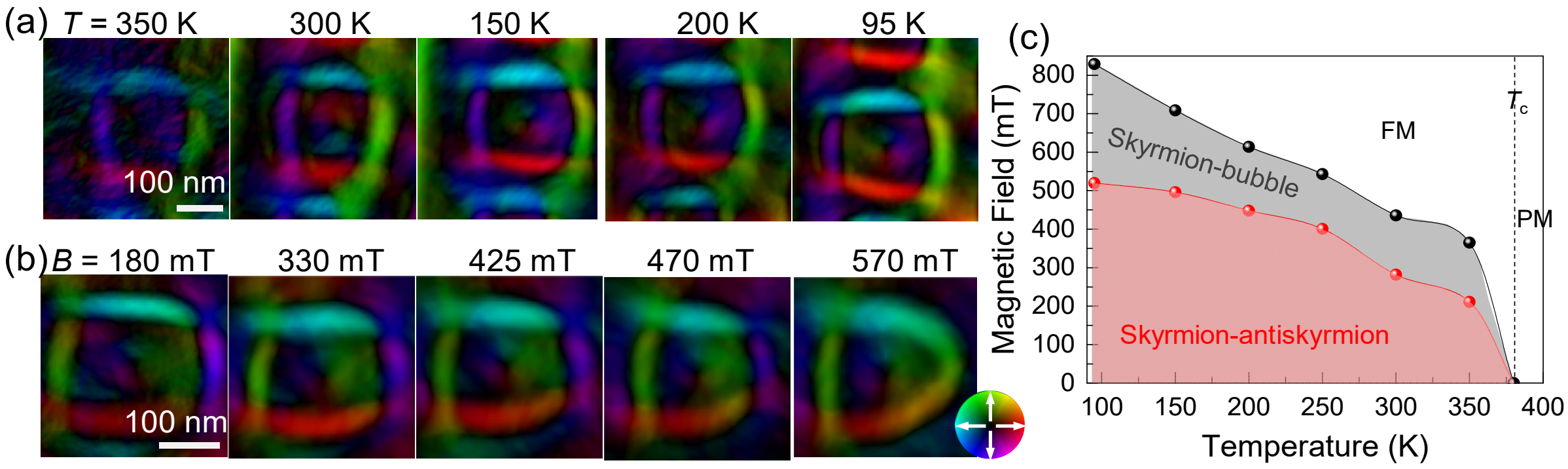

Fig. S19. (a) Skyrmion-antiskyrmion strings survive as metastable phases in a field-cooling process at $B$ = 93 mT. Defocused distance, 300 μm. (b) Transformation from the skyrmion-antiskyrmion string to the skyrmion-bubble string in $B$-increasing process at $T$ = 95 K. (c) Phase diagram of the skyrmion-antiskyrmion string as the metastable state. The dots represent threshold magnetic fields between different phases.

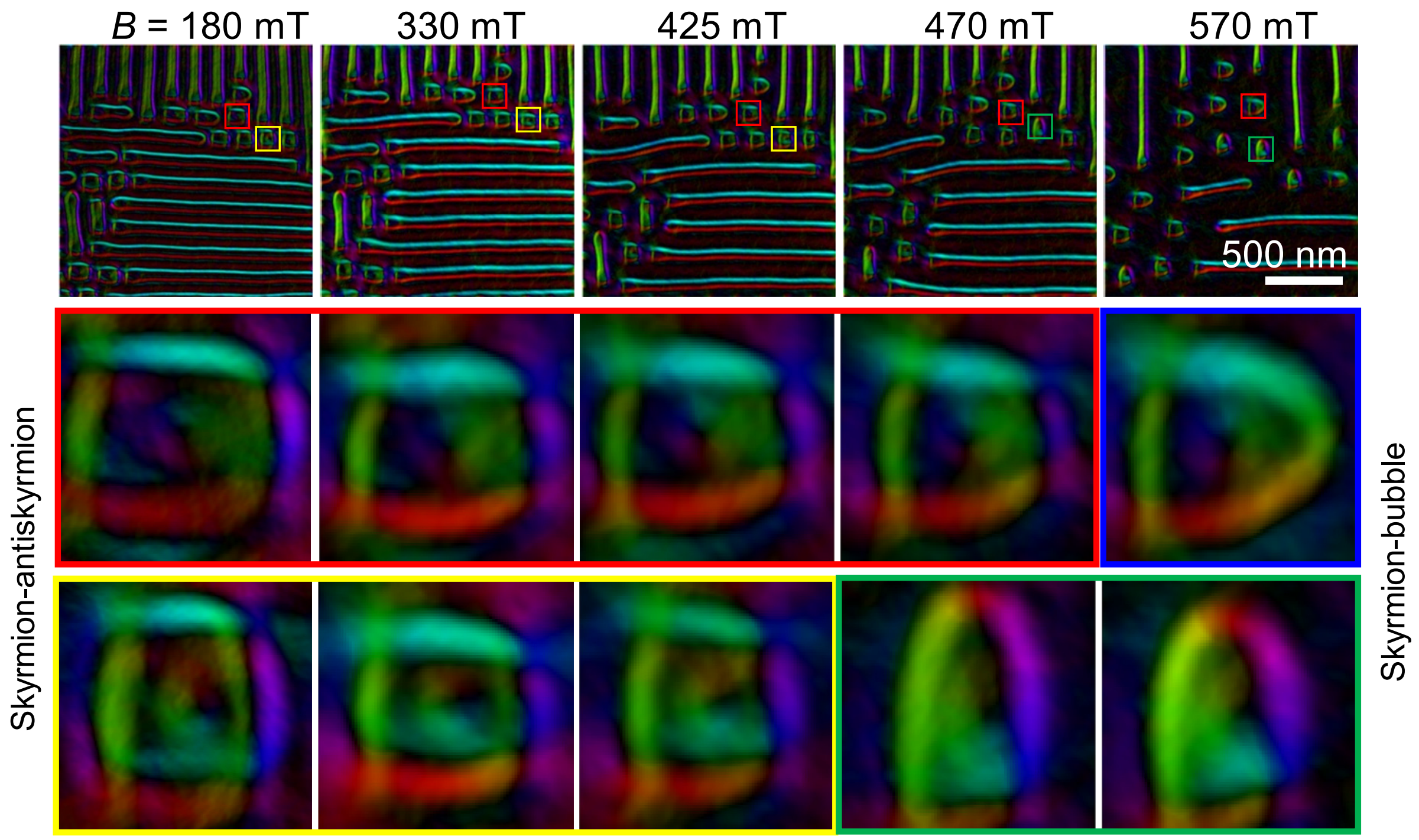

Fig. S20. Transformation from hybrid skyrmion-antiskyrmion strings to skyrmion-bubble strings in the field-increasing process obtained from TIE analysis. Defocused distance, 300 μm. Temperature $T$ = 95K.

**Supplementary Section II**

**Dynamics of the skyrmion-antiskyrmion string**

Skyrmions with an integer $Q$ experience sideway motions perpendicular to the spin-polarised current, which is known as the skyrmion Hall effect [5, 6]. Zero skyrmion Hall effects have been theoretically proposed to be realized by $Q$ = 0 skyrmion bundles [7], skyrmion-antiskyrmion coupling pairs [8, 9], and antiferromagnetic skyrmions [10, 11]. The hybrid skyrmion-antiskyrmion tube with greatly decreased averaged $Q$ (Supplemental Figure S2h) could provide a new strategy to realize Hall balance. We thus further explore the current-induced dynamics of the skyrmion-antiskyrmion string in simulation, as shown in Figure S21. Simulations reveal the topology-dependent dynamic motions for the three types of skyrmionic textures (Figure S21a). Excitingly, the skyrmion-antiskyrmion string moves only along the current orientation with negligible skyrmion Hall effects. The average $Q$ of the skyrmion string, antiskyrmion string, and skyrmion-antiskyrmion string are $-1$, 1, and $-0.3$, respectively. However, the simulated Hall angles are not simply proportional to $Q$. The Hall angle $\theta_h$ derived from the rigid approximation of Thile's equation is expressed by [7]:

$$\theta_h = \arctan\left[\frac{(\alpha-\beta)\eta_{xx}Q}{Q^2+(\eta_{xx}\eta_{yy}-\eta_{xy}\eta_{yx})\alpha\beta-\eta_{xy}Q(\alpha-\beta)}\right] \tag{S3}$$

Here, $\alpha$ is the Gilbert damping, and $\beta$ is the non-adiabatic parameter. Based on equation (S3), the Hall angle $\theta_h$ is dependent on not only $Q$ but also the shape factor $\eta_{ij}$, a matrix with each term written as $\eta_{i,j} = (1/4\pi)\int(\partial_i\mathbf{m}\cdot\partial_j\mathbf{m})\,\mathrm{d}x\mathrm{d}y$. Here, $\mathbf{m}$ is the unit magnetization. For the twisted skyrmion-antiskyrmion string, the shape factor varies with the depth $z$ (Figure S21b). Thus, the theoretical Hall angle also varies with depth for these skyrmionic textures (Figure S2c). We identify that the theoretical Hall angles averaged over the thickness for the skyrmion string, the antiskyrmion string, and the skyrmion-antiskyrmion string are 18.3°, 32.7°, and 2.6° respectively. These theoretical average Hall angles agree well with the simulated results in Figure S21a. Finally, we show that the negligible skyrmion Hall effect of the skyrmion-antiskyrmion string works for a broad region of material parameters ($\alpha$ and $\beta$), as shown in Figure S21d.

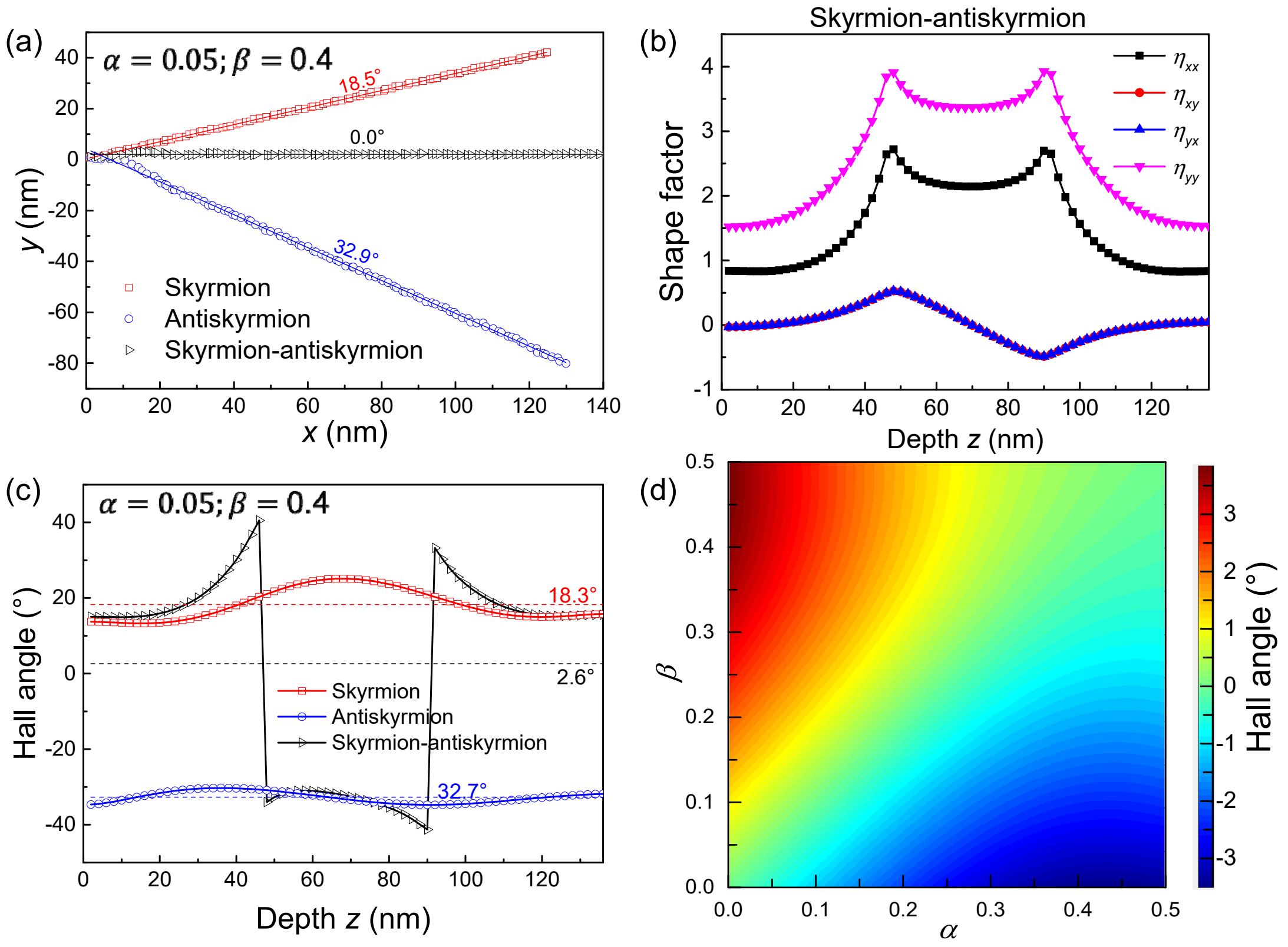

**Fig. S21.** Simulated Hall balance of the skyrmion-antiskyrmion string. (a) Trajectories of three types of skyrmionic configurations driven by 10-ns currents with a density of $10^{11}$ A m$^{-2}$. The current is applied along the $x$-axis. (b) Depth $z$ dependence of shape factor of the 3D skyrmion-antiskyrmion string. (c) Theoretical skyrmion Hall angle of three types of skyrmionic configurations. The dash lines marked the average Hall angle over the depth. (d) Theoretical skyrmion Hall angle of the skyrmion-antiskyrmion string as a function of Gilbert damping $\alpha$ and non-adiabatic parameter $\beta$. Two-dimensional periodical condition boundary conditions were used. Magnetic field $B/(0.5\mu_0 M_s) = 1.45$.